\documentclass[12pt]{article}
\usepackage{graphicx,euscript,psfrag,epsf}

\newcommand{\LQCD}{\Lambda_\mathrm{QCD}}

\newcommand{\be}{\begin{equation}}
\newcommand{\ee}{\end{equation}}
\newcommand{\bea}{\begin{eqnarray}}
\newcommand{\eea}{\end{eqnarray}}
\newcommand{\nl}{\nonumber \\}

\newcommand{\ov}[1]{ \overleftarrow{#1} }

\def\slash#1{#1 \hskip-0.45em /}
\def\Slash#1{#1 \hskip-0.59em /}

\makeatletter
\def\fmslash{\@ifnextchar[{\fmsl@sh}{\fmsl@sh[0mu]}}
\def\fmsl@sh[#1]#2{%
  \mathchoice
    {\@fmsl@sh\displaystyle{#1}{#2}}%
    {\@fmsl@sh\textstyle{#1}{#2}}%
    {\@fmsl@sh\scriptstyle{#1}{#2}}%
    {\@fmsl@sh\scriptscriptstyle{#1}{#2}}}
\def\@fmsl@sh#1#2#3{\m@th\ooalign{$\hfil#1\mkern#2/\hfil$\crcr$#1#3$}}
\makeatother
\def\beq{\begin{eqnarray}}
\def\eeq{\end{eqnarray}}

\def\eps{\epsilon}

\def\be{\begin{equation}}
\def\ee{\end{equation}}

\def\np{n_+}
\def\nm{n_-}

\def\qslash{\rlap{\hspace{0.02cm}/}{q}}

\def\LQCD{\Lambda_{\rm QCD}}

\def\J{{\EuScript J}}

\def\J{{\EuScript J}}

\begin{document}

\renewcommand{\thefootnote}{\fnsymbol{footnote}}
\begin{titlepage}

\begin{flushright}
PITHA 04/18\\
SI-HEP-2004-12\\
SFB/CPP-04-64\\
hep-ph/0411395\\[0.0cm]
22 November 2004
\end{flushright}

\vspace{1cm}
\begin{center}
\Large\bf\boldmath
Power corrections to $\bar{B}\to X_u \ell  \bar{\nu}$ ($X_s\gamma$) 
\unboldmath decay spectra in the
``shape-function'' region 
\end{center}

\vspace{0.5cm}
\begin{center}
M.~Beneke${}^{\,(a)}$, 
F.~Campanario${}^{\,(b)}$\footnote{Address after December 1, 2004:
Department of Physics, University of Toronto,
60~St.~George Street, Toronto, Ontario, Canada M5S 1A7}, 
T.~Mannel${}^{\,(c)}$, 
B.D.~Pecjak${}^{\,(c)}$\\[0.3cm]
{\sl ${}^{(a)\,}$Institut f\"ur Theoretische Physik E, RWTH Aachen, 
D -- 52056 Aachen, Germany\\
${}^{(b)\,}$Institut f\"ur Theoretische Teilchenphysik, Universit\"at
Karlsruhe, D -- 76128 Karlsruhe\\
${}^{(c)\,}$Theoretische Physik 1, Fachbereich Physik, Universit\"at Siegen,
D -- 57068 Siegen, Germany
}
\end{center}

\vspace{0.8cm}
\begin{abstract}
\vspace{0.2cm}\noindent 
Using soft-collinear effective theory (SCET), 
we examine the $1/m_b$ corrections 
to the factorization formulas for inclusive semi-leptonic $B$ decays in the 
endpoint region, where the hadronic final state consists of a single 
jet. At tree level, we find a new contribution from four-quark 
operators that was previously assumed absent.  Beyond tree level many 
sub-leading shape-functions  are needed to correctly describe
the decay process.
\end{abstract}

\end{titlepage}

\renewcommand{\thefootnote}{\arabic{footnote}}
\setcounter{footnote}{0}
\section{Introduction}

Inclusive semi-leptonic and radiative $B$ meson decays 
offer many opportunities to test the flavour sector of the Standard Model. 
The radiative decay $\bar{B}\to X_s\gamma$ can be used
as a probe for new physics, and the most precise measurements 
of the CKM matrix element $|V_{ub}|$ are based on the semi-leptonic decay
$\bar{B}\to X_u \ell\bar\nu$. However, experimental cuts on the 
final state are needed to suppress a strong background 
from charm production.  These cuts often force the kinematics of the 
decay into the so-called shape-function region, where the hadronic 
final state is collimated into a single jet, which carries a large energy
of order $m_b$, but a small invariant mass of 
order $m_b\Lambda_{\rm QCD}$.  While the conventional operator product 
expansion (OPE) breaks down in this region of phase-space, it is still possible
to make rigorous predictions using a twist expansion, which sums singular terms
in the OPE into  non-local operators evaluated on the light-cone 
\cite{Neubert:1993ch, Bigi:1993ex, Mannel:1994pm}.  This method amounts to 
using QCD factorization formulas which separate the physics from the
disparate mass scales $m_b^2 \gg m_b \LQCD \gg \LQCD^2$ into hard, jet-, and
shape-functions, respectively. To the above mentioned decays these ideas 
have been applied in \cite{Korchemsky:1994jb,Akhoury:1995fp}.  
The recent development of the 
soft-collinear effective theory (SCET) as applied to inclusive decays 
\cite{Bauer:2000yr, Bauer:2001yt, Beneke:2002ph,Beneke:2002ni} has provided a 
natural framework from which to prove these QCD factorization 
formulas to all orders in perturbation theory, and also to sum 
the Sudakov logarithms appearing 
therein; such studies have been performed to leading order in
$1/m_b$ \cite{Bauer:2001yt, Bosch:2004th, Bauer:2003pi}.  

The goal of the present work is to use SCET to analyze the structure
of the  factorization formulas beyond leading order in $1/m_b$.  
The motivation for this is two-fold.  
First, from the phenomenological side, the inclusion of power-suppressed
effects will soon be necessary to keep theoretical predictions 
on par with the improving experimental accuracy being achieved at
 the $B$ factories.  
Second, from a purely theoretical standpoint, 
although the machinery needed to begin a discussion of
 power corrections to QCD factorization formulas within SCET has
been available for some time, there have been no theoretical efforts
which apply this to a concrete example. For inclusive decays a study of 
sub-leading shape-function effects was carried out previously in  
\cite{BLM,Bauer:2002yu,Leibovich:2002ys,Burrell:2003cf} by more 
elementary methods, and restricted to the tree 
approximation. In this paper, we take the first steps towards analyzing 
power-suppressed effects in inclusive $B$ decays in the shape-function
region beyond this approximation by an investigation of the general 
structure of factorization, and an enumeration of all relevant 
shape-functions. We perform a complete calculation of the hadronic 
tensor only at tree level. However, even in 
this approximation we find a new effect related to certain four-quark 
operators, and some discrepancies with earlier results.

The factorization formulas are derived by performing a two-step 
matching from  ${\rm QCD \to SCET \to HQET}$, identifying the 
hard functions in the first step of matching, and the jet-  
and shape-functions in the second. The first matching is done at 
scales of order $m_b$, the second at scales $\sqrt{m_b\LQCD}$. 
Since the hard and jet-functions can therefore be computed perturbatively 
in the strong coupling, it is evident that factorization extends 
to sub-leading order in $1/m_b$. The advantage of the effective 
theory framework is that it provides a transparent book-keeping 
of all the relevant interactions at every step in the calculation, 
including power-suppressed effects, and allows us to  
identify all the HQET operators that define the sub-leading 
shape-functions with no restriction to tree level in the 
hard ($\mu\sim m_b$) 
and hard-collinear ($\mu\sim\sqrt{m_b\LQCD}$) fluctuations.

The content of the paper is as follows: 
Sections~\ref{sec:fact} to \ref{sec:basis} provide an outline of 
the structure of factorization at sub-leading order in $1/m_b$,
including a basis for the SCET currents and HQET shape-functions 
needed at order $1/m_b$, but to any order in the strong coupling 
$\alpha_s$. In Section~\ref{sec:example} we pick one of the 
many terms in the expansion and demonstrate its reduction to a
convolution of a jet- and a shape-function. In 
Section~\ref{sec:tree} we compute the structure functions that
parameterize the semi-leptonic and radiative decay spectra 
in the tree approximation at order $1/m_b$. We find many 
structural simplifications at tree level, but also a tetra-local 
term that has been missed before. Sections~\ref{sec:beyondtree} and 
\ref{sec:phenom} follow up on these observations 
by commenting on the nature of possible simplifications beyond tree
level and by providing a numerical  estimate of the correction 
from the new tetra-local contribution. We conclude in 
Section~\ref{sec:conclude}.

\section{Factorization at sub-leading order }
\label{sec:fact}

The effects of QCD in semi-leptonic $B$ decays are contained
in the hadronic tensor $W^{\mu\nu}$.  We can calculate the 
hadronic tensor via the optical theorem by taking the imaginary
part of the forward scattering amplitude, which we define
as 
\be
W^{\mu\nu}= \frac{1}{\pi}\,{\rm Im}\,\langle\bar{B}(v)
|T^{\mu\nu}|\bar{B}(v)\rangle.
\label{hadronic_tensor}
\ee
We use the state normalization $\langle\bar{B}(v)|\bar{B}(v)\rangle$=1
and drop the velocity label from now on. 
The  correlator $T^{\mu\nu}$ is the time-ordered product 
of two flavour-changing weak currents
\be \label{eq:correlator}
T^{\mu\nu}=i\int d^4 x e^{-iq\cdot x}{\rm
  T}\{J^{\dagger\mu}(x),J^{\nu}(0)\},
\ee
where $q$ is the momentum carried by the outgoing leptonic pair
($\bar B\to X_u \ell\bar \nu$) or photon ($\bar B\to X_s\gamma$). 
The calculation of this correlator is the essential
element to obtaining the decay amplitudes. Any differential decay 
distribution can be derived from the components of the hadronic 
tensor. We shall use the notation of \cite{NeubertDeFazio} and 
write the hadronic tensor in terms of five scalar structure functions  
\begin{eqnarray} 
W_{\mu \nu} &=& W_1\,  (p_\mu v_\nu + v_\mu p_\nu - g_{\mu \nu} vp 
- i \epsilon_{\mu \nu \alpha \beta} \,p^\alpha v^\beta ) \\ \nonumber 
&& - \,W_2 \, g_{\mu \nu} + W_3 \, v_\mu v_\nu 
+ W_4 \, (p_\mu v_\nu + v_\mu p_\nu) + W_5 \, p_\mu p_\nu ,
\label{hadtensordef}
\end{eqnarray} 
where the independent vectors are chosen to be $v$, the velocity of 
the $\bar B$ meson, and $p\equiv m_b v-q$ 
with $m_b$ the $b$ quark pole mass. (We use the convention 
$\epsilon^{0123}=-1$.) 
The $W_i$ are regarded as functions of $p^2$ and $vp$. The relation 
of these partonic variables  
to the final state hadronic invariant mass $P^2$ and energy $v P$ is 
\begin{equation}
\label{hadvar}
vp=v P -(M_B-m_b),\qquad p^2= P^2 - 2(M_B-m_b)v P+ (M_B-m_b)^2.
\end{equation}
Of course, there is a dependence of the $W_i$ on the current $J$ assumed in 
(\ref{eq:correlator}), which is not made explicit in our notation.
Since we are interested in either the semi-leptonic or radiative
decay, the relevant currents are\footnote{For radiative 
decays the restriction to currents implies that we consider only 
the contribution from the leading electromagnetic penguin operator 
in the effective weak Hamiltonian. The complete result contains 
additional terms from four-quark operators.}
\begin{equation}
\label{eq:QCDcurrents}
J^\mu=\bar{q}\gamma^\mu(1-\gamma_5)b, \qquad
J^\mu=\frac{1}{2}\,\bar{q}\,[\gamma^\mu, \qslash\,](1+\gamma_5)b.
\end{equation}

The decay kinematics are assumed to be in the so-called shape-function or
SCET region, where the hadronic jet emitted at the weak vertex carries
a large energy of order $m_b$, but a small invariant mass squared of order
$m_b \Lambda_{\rm QCD}$.  Such a jet is referred to as 
hard-collinear,
and implies the existence of three
widely separated mass scales $m_b^2  \gg  m_b \LQCD \gg \LQCD^2$,  
along with a small parameter $\lambda^2=\LQCD/m_b$. 
The soft-collinear effective theory (SCET) offers a 
tool with which to calculate the decay amplitudes as 
a series in this small parameter $\lambda$. 
An arbitrary momentum $p$ is decomposed as
\be
p^\mu=\np p\,\frac{\nm^\mu}{2}+ p_\perp^\mu + \nm p\,\frac{\np^\mu}{2},
\ee
where $n_{\pm}^\mu$ are light-like vectors satisfying $\np\nm=2$.  
For inclusive decays it is sufficient to use what is referred to 
in the literature as $\rm{SCET_I}$, which contains only hard-collinear
and soft degrees of freedom. The components of a hard-collinear 
momentum are defined to scale 
as 
$(\np p_{hc}, \,p_{hc\perp}, \,\nm p_{hc})
\sim m_b(1,\,\lambda,\,\lambda^2)$ 
and those of a soft momentum as 
$p_s \sim m_b(\lambda^2,\,\lambda^2,\,\lambda^2)$. From now on
we will refer to hard-collinear momenta as simply collinear,
and to ${\rm SCET_I}$ as SCET. We find it convenient to work in 
a frame of reference where $v_\perp=0$ and $n_- v = n_+ v=1$. 
Furthermore, the SCET expansion in $\lambda$ refers to a frame 
in which the total transverse momentum of the hadronic final state 
is at most of order $\lambda$. For any given $q$, we therefore choose
the frame where $q_\perp=[m_b v-p]_\perp = -p_\perp=0$ and 
compute the invariant 
components of the hadronic tensor in this frame.

It has been shown that the decay amplitude factorizes into
a convolution of hard, jet, and soft functions at leading order in
$\lambda$ \cite{Korchemsky:1994jb,Bauer:2001yt}. 
Using SCET, our aim is to show that an analogous 
factorization holds at sub-leading order in $\lambda$. 
The scaling properties of the collinear jet imply that no Lorentz
invariant quantity can be formed at order $\lambda$, so the leading
power corrections appear at order $\lambda^2$. In technical terms, 
transverse momenta of order $\lambda$ appear only in the internal 
integrations over the collinear momenta in a jet, and since the 
integral over an odd number of transverse momenta either vanishes 
or must be proportional to one of the external soft transverse 
momenta of order $\lambda^2$, the resultant expansion is in 
powers of $\lambda^2\sim 1/m_b$. 
We therefore need to go to second order 
in the SCET expansion. 
Before giving any explicit formulae, we will outline a 
procedure that as a matter of principle could 
be used to establish factorization at any order in $\lambda$.

\paragraph{\it 1. Matching to SCET/HQET.} In the first step we 
remove the hard scale $m_b^2$ as a dynamical scale 
by matching the QCD Lagrangian and currents
onto their corresponding expressions in HQET and SCET, where fluctuations 
are characterized by the jet scale $m_b\LQCD$.  We then use these 
effective theory quantities to calculate the correlator in 
(\ref{eq:correlator}). In the following we use the position 
space formulation of SCET \cite{Beneke:2002ph,Beneke:2002ni}, which is
especially well suited for the study of power corrections, since the 
power-suppressed Lagrangians and currents are already known. 

We first discuss the SCET/HQET Lagrangian. 
Its explicit form to order $\lambda^2$ is \cite{Beneke:2002ni}
\begin{eqnarray}
{\cal L}&=&\bar{\xi}\left(i \nm D +i\Slash{D}_{\perp c}\frac{1}{i\np D_c}
i\Slash{D}_{\perp c}\right)\frac{\slash{n}_+}{2}\xi 
 -\frac{1}{2}\,\mbox{tr}\left(F_c^{\mu\nu}
  F^c_{\mu\nu}\right) 
\nonumber\\
&&+\, \bar h_v i v D_s h_v + \bar q_s i\Slash{D}_{s} q_s 
 - \frac{1}{2}\,\mbox{tr}\left(F_s^{\mu\nu} F^s_{\mu\nu}\right) 
\nonumber\\
&&+\, {\cal L}^{(1)}_{\xi}+{\cal L}^{(1)}_{\xi q}+ {\cal L}^{(1)}_{\rm YM}
+ \sum_{i=1}^3 {\cal L}^{(2)}_{\xi i}+{\cal L}^{(2)}_{\xi q}
+ {\cal L}^{(2)}_{\rm YM} + {\cal L}_{\rm HQET}^{(2)}.  
\label{lagrangian}
\end{eqnarray}
The $\xi$ denotes the collinear quark field, $q_s$ the soft quark field, and 
$h_v$ the heavy quark field of HQET. The covariant derivatives 
are defined as $iD_c = i\partial + g A_c$ and analogously for $iD_s$, 
but a $D$ without subscript contains both the collinear and soft gluon 
field. The quantities $F_{\mu\nu}^s$ ($F_{\mu\nu}^c$) are 
the field-strength tensors built from the soft (collinear) gauge 
fields in the 
usual way, except for the definition of $F_{\mu\nu}^c$, where 
$n_- D$ rather than $n_- D_c$ appears.  The collinear and soft 
fields are evaluated at $x$, but in products of soft and 
collinear fields the soft fields are evaluated  
at $x_-^\mu=(\np x/2)\,\nm\equiv x_+\nm^\mu$, according to the multipole
expansion.  In this notation $x_+$ is a 
scalar, while $x_-^\mu$ is a vector.  

The power-suppressed terms in the effective Lagrangian read 
\bea
\label{eq:lagrangians}
{\cal L}_\xi^{(1)}&=&\bar{\xi}x_{\perp}^\mu n_-^\nu W_c \,g F^s_{\mu\nu} W_c^\dagger
\frac{\slash{n}_+}{2}\xi, 
\nonumber \\
{\cal L}^{(2)}_{1\xi } & = & \frac12\bar{\xi}\nm x n_+^\mu n_-^\nu 
W_c \,g F^s_{\mu\nu} W_c^\dagger\frac{\slash{n}_+}{2}
\xi, 
\nonumber \\
 {\cal L}^{(2)}_{2\xi } &=& \frac12\bar{\xi}x_\perp^\mu x_{\perp\rho}n_-^\nu
W_c[D_{\perp s}^{\rho},g
F_{\mu\nu}^s]W_c^\dagger\frac{\slash{n}_+}{2}\xi,
\nonumber \\
{\cal L}^{(2)}_{3\xi } &=&\frac{1}{2} \bar{\xi}  i\Slash{D}_{\perp c}
\frac{1}{i\np D_c}
x_{\perp}^\mu \gamma_{\perp}^\nu W_c \,g F^s_{\mu\nu}W_c^\dagger
\frac{\slash{n}_+}{2}\xi \nonumber \\
&&+\,\frac12\bar{\xi}x_\perp^\mu\gamma_{\perp}^{\nu}
W_c \,g F^s_{\mu\nu}W_c^\dagger\frac{1}{i\np D_c}
i\Slash{D}_{\perp c}\frac{\slash{n}_+}{2}\xi, 
\nonumber \\
{\cal L}^{(1)}_{\xi q}&=&\bar q_s W_c^\dagger i\Slash{D}_{\perp c}\xi - 
\bar \xi i\ov{\Slash D}_{\perp c} W_c q_s, 
\label{eq:Lqxi}
\eea
where the $W_c$ are collinear Wilson lines. We have omitted the 
terms ${\cal L}^{(2)}_{\xi q}$, since their field content 
implies that they do not 
contribute to the current correlator at order $\lambda^2$. 
The explicit expressions of the Yang-Mills Lagrangians 
${\cal L}^{(1,2)}_{\rm YM}$ can be found in \cite{Beneke:2002ni}. 
They are needed only in the calculation of $1/m_b$ corrections 
beyond tree level. The SCET Lagrangian is exact to all orders in 
perturbation theory, receiving no radiative 
corrections \cite{Beneke:2002ph}. In contrast, the 
$\lambda^2$ HQET Lagrangian is 
\be
\label{lhqet}
{\cal L}^{(2)}_{\rm HQET}=\frac{1}{2 m_b}\left[\bar h_v (iD_s)^2 h_v 
+ \frac{C_{\rm mag}}{2}\,\bar h_v\sigma_{\mu\nu}g F_s^{\mu\nu}h_v\right],
\ee 
where $C_{\rm mag}\not= 1$ 
represents the renormalization of the chromomagnetic 
interaction by hard quantum fluctuations. 

A second source of hard 
corrections is related to the matching onto
the SCET heavy-light currents. The QCD 
currents $J_i = \bar{\psi} \hspace*{0.03cm}\Gamma_i  \hspace*{0.03cm}Q$ 
are represented in SCET as convolutions of dimensionless 
short-distance Wilson coefficients depending on quantities at
the hard scale $m_b^2$ with current operators  
$J^{(k)}_j$ composed of HQET and SCET fields. The matrix elements 
of these effective currents are  
characterized by fluctuations on the order of the jet
scale $m_b \Lambda_{\rm QCD}$ and the soft scale $\LQCD^2$.    
We write this convolution
as 
\begin{eqnarray}
\label{match}
(\bar{\psi}\hspace*{0.03cm}\Gamma_i \hspace*{0.03cm}Q)(x)&=& 
e^{-i m_b v\cdot x}\,
 \sum\limits_{j,k} \widetilde{C}^{(k)}_{ij}(\hat{s}_1,
\ldots,\hat{s}_n)\otimes
J_{j}^{(k)}(\hat{s}_1,\ldots,\hat{s}_n;x),
\end{eqnarray}
where the $\otimes$ stands for a convolution over 
a set of dimensionless variables $\hat{s}_i\equiv s_i m_b$. 
The superscript $k$
refers to the scaling of the current operator with $\lambda$ relative 
to the leading-power currents, 
and the subscript $j$ enumerates the effective currents at a given order 
in $\lambda$. 
As with the SCET Lagrangian, the collinear fields in the current
operator $J_j^{(k)}$
depend on $x+s_i n_+$ but the soft fields including $h_v$ are 
multipole-expanded and depend only on $x_-$. 
The factorization formula will eventually be
formulated in terms of convolutions over longitudinal collinear 
momentum fractions,
which we define as  $u_i=\np p_i/m_b$.
The momentum space
coefficient functions are related to those defined above by 
\begin{equation}\label{eq:umatch}
  C^{(k)}_{ij}(u_i)=\int \prod \limits_i d\hat{s}_i\,
  \widetilde{C}^{(k)}_{ij}(\hat{s}_i)  \,e^{i  \sum \limits_i \hat{s}_i
u_i}.
\end{equation}
A basis for the most general set  
of order $\lambda^2$ currents including radiative corrections 
has not yet been discussed in the literature. We shall return to this 
point in Section \ref{sec:currents}. 
For now, we simply note that these short-distance 
Wilson coefficients depend only on quantities at the
hard scale $m_b^2$ and are identified with the hard functions 
in the factorization formula.

We have now achieved the factorization of the hard scale from 
the soft and collinear degrees of freedom 
and can write the correlator as  
\be
T^{\mu\nu}=\tilde H_{jj'}(\hat s_1,\dots, \hat s_n)\otimes 
T^{{\rm eff},\mu\nu}_{jj'}(\hat s_1,\dots, \hat s_n),
\ee
The hard function  $\tilde H_{jj'}$ is a product of 
SCET Wilson coefficients
$\tilde C_{ij}^{(k)} \tilde C_{i'j'}^{(k')}$ (along with a possible 
$C_{\rm mag}$ from the HQET Lagrangian) and 
$T^{{\rm eff},\mu\nu}_{jj'}$ is a correlator of 
SCET currents. In what follows we will suppress all indices and denote
the correlator by $T^{{\rm eff}}$.  

\paragraph{\it 2. Collinear-soft factorization.} 
In the second step we factorize the matrix element of 
$T^{\rm eff}$ into
soft and collinear pieces. 
Towards this end, we first redefine  the collinear fields 
according to \cite{Bauer:2001yt}
\be\label{eq:redef}
\xi = Y\xi^{(0)},\quad A_c =  
Y A_c^{(0)} Y^\dagger, \quad W_c =  Y W^{(0)}_c Y^\dagger, 
\ee
and immediately drop the superscript on the redefined fields. 
The $Y$ above is a soft Wilson line involving $n_- A_s$ evaluated at $x_-$, 
and the redefinition  of the collinear Wilson line follows from
that of the gluon field. The effect of this redefinition on 
the SCET Lagrangian and currents is to transform every 
$n_- D$ into $n_- D_c$, and to replace $q_s$ ($h_v$) by 
$Y^\dagger q_s$ ($Y^\dagger h_v$) and $i D^s_\mu$ ($F_{\mu\nu}^s$) 
by $Y^\dagger i D^s_\mu Y$ ($Y^\dagger F_{\mu\nu}^s Y$). Note that 
the collinear Wilson line appears only in $W_c^\dagger \xi$ and 
$W_c^\dagger  iD^c_\mu W_c$, so the positions of soft and collinear 
Wilson lines in 
any field product can always be inferred from the transformation of 
the fields under the collinear and soft gauge symmetries. 

We shall treat the power-suppressed Lagrangian terms in 
the interaction picture. Since after the collinear field redefinition 
the leading order SCET Lagrangian (see (\ref{lagrangian})) 
no longer couples the collinear to the soft fields, we can 
factorize the matrix element. The $\bar B$ meson by definition 
contains no collinear degrees of freedom, meaning that 
the $\bar B$ meson state is represented as the tensor product 
$|\bar B\rangle\otimes |0\rangle$, where the first (second) factor 
refers to the soft (collinear) Hilbert space. It follows that 
the matrix element of any SCET current correlator, including in general 
time-ordered products with sub-leading interactions from the 
Lagrangian, can be written as  
\bea
&&\langle\bar B| T^{\rm eff}(\hat s_1,\dots,\hat s_n)|\bar B\rangle=
i\int d^4 x d^4 y\dots e^{i(m_b v-q) x}\langle\bar{B}
|\bar{h}_v[{\rm soft\, fields}]h_v|\bar{B}\rangle(x_-,y_-,...)
\nonumber\\
&& \hspace*{2cm}\times\, 
\langle 0 |[{\rm collinear\, fields}] |0\rangle(\hat s_1,\dots,\hat s_n;
x,y,...). 
\eea
The additional integrals over $d^4 y\dots$ are related to 
insertions of the power suppressed Lagrangian. The soft matrix element 
depends only on $n_+ x, n_+ y,\ldots$, so the integrations 
over transverse positions and the $n_-$ components can be lumped 
into the definition of the collinear factor. The soft and 
collinear matrix elements are then  linked by multiple convolutions 
over the light-cone variables $x_+, y_+,\ldots$.

\paragraph{\it 3. Definition of ``shape-functions'' 
and ``jet-functions''.} 

The matrix element of the soft fields between $\bar B$ meson
states defines non-perturbative (leading and sub-leading) shape-functions
$\tilde S(\np x,\dots)$.  We define these in momentum space
according to
\be
 \tilde S(x_{1+},\dots, x_{n+})= \int d \omega_1 \dots 
d\omega_n\,
e^{-i( \omega_1 x_{1+} \dots + \omega_n x_{n+})}\,
 S(\omega_1,\dots,\omega_n).
\label{defshape}
\ee
We shall see below that at order 
$\lambda^2$ we can have up to a triple convolution over
the variables $\omega_i$, coming from two insertions of the
order $\lambda$ SCET Lagrangian.     
These scalar functions need to be modeled and introduce hadronic
uncertainties into any phenomenological applications.  
Properly identifying the independent functions is therefore
an important task to which we devote Sections \ref{sec:basis}
and \ref{sec:beyondtree}. 

The matrix elements of the collinear fields 
between the collinear 
vacuum involve fluctuations with virtualities $m_b \LQCD$. 
After integration over position arguments they 
define perturbatively calculable jet-functions, which 
have the form 
\be
\J(u_1, \dots, u_i;\omega_1, \dots \omega_n).
\ee
(Alternatively, we shall use the variables $p^2_{\omega_{12\dots j}}$, 
where $p_{\omega_{12\dots j}}=p-\omega_{12\dots j} \,\np/2$ and 
$\omega_{12\dots j}=\omega_1
+\omega_2+\dots \omega_j$.)      
Evaluating
these functions removes the collinear degrees of freedom
and defines the final step of matching, ${\rm SCET}\to {\rm HQET}$.
The complicated form of the power-suppressed SCET currents and 
Lagrangian imply that many new jet-functions appear at order 
$\lambda^2$ and we will not list the general set in this paper.
Instead, we simply note that they are  perturbative objects
and so introduce no inherent theoretical uncertainties.  

Having carried out these steps, we can  express the correlator in terms 
of a factorization formula. We insert the definitions of the 
hard coefficients, jet-functions,
and shape-functions and work in momentum space.  A generic 
term in the factorization formula is a sum of convolutions over
hard, jet, and shape-functions 
\be
 T=\sum H( u_1,\dots u_i)\otimes 
\J(u_1,\dots,u_i;\omega_1, \dots, \omega_n)\otimes S(\omega_1,\dots,\omega_n)
\label{eq:masfact}.
\ee
To discuss more precisely the structure of this
factorization formula, we need to identify the set of 
jet- and shape-functions appearing at order $\lambda^2$.  
For the jet-functions, the necessary step is to derive the
full set of heavy-light currents in the presence of
radiative corrections.  We turn to this topic in the next
section.  

\paragraph{\it 4. Caveats to factorization.} 

The operators whose matrix elements define the jet- and shape-functions 
have singularities related to the light-cone expansion. Above it 
has been assumed that there exists a regularization 
and subtraction procedure that is compatible with the properties 
of the Lagrangian crucial to factorization. To the best of our knowledge, 
dimensional regularization is adequate for this purpose, but we do 
not have a general proof of this statement. 

The factorization formula (\ref{eq:masfact}) is composed of convolutions 
over the soft light-cone variables $\omega_i$, and we must assume 
that the convolutions of the perturbative jet-functions with the 
shape-functions converge. Little is known about the functional dependence 
of sub-leading shape-functions, but a divergence of the convolution 
for $\omega_i\to 0$ would be surprising, since it would indicate that  
long-distance physics is not accounted for by the HQET Lagrangian. On the 
other hand, divergences for $\omega_i\to\infty$, if they existed, 
would be of short-distance nature and could presumably be treated 
with a modification of the factorization procedure.

\section{SCET currents}
\label{sec:currents}
In this section we discuss the matching of the QCD heavy-light
currents onto SCET. The order 
$\lambda$ operators have been investigated in several 
places \cite{Beneke:2002ph, Beneke:2002ni, Pirjol:2002km}  
including one-loop radiative corrections 
to their coefficients \cite{Beneke:2004rc,Hill:2004if, Becher:2004kk}, but 
the order $\lambda^2$ case has been given only at 
tree level \cite{Beneke:2002ph, Beneke:2002ni}.  
To give  a general factorization formula for the 
correlator, we will need a complete basis for the  
heavy-light currents in the presence of radiative corrections.  

The QCD currents $J_i = \bar{\psi} \hspace*{0.03cm}\Gamma_i 
\hspace*{0.03cm}Q$ 
are represented in SCET as a convolution of dimensionless 
short-distance Wilson coefficients with a current operator 
$J^{(k)}_j$
composed of heavy-quark and SCET fields, as in (\ref{match}).
Our task is to find  a set of  current operators $J_j^{(k)}$ up to 
order $\lambda^2$ relative to the leading-order currents.  
We work in the frame where $v_\perp=0$,
which considerably reduces the number of allowed currents at a given
order in $\lambda$.  

In currents at position $x$, all soft fields are evaluated at 
$x_-= (\np x/2) \nm$, but the collinear fields may be shifted 
along the light-cone to positions  $x+s_i n_+$ owing to the 
non-locality of SCET. The variables 
$s_i$ are integrated in a convolution product of coefficient 
functions and current operators, which allows us to 
eliminate factors of $i n_+ \partial$ and 
$1/(i n_+\partial)$ from the operator 
by a redefinition of the coefficient function. In the following 
we use the convention that operators do not 
contain $i n_+\partial$ (or its inverse) operating on collinear fields, and 
denote the position argument $x+s_i n_+$ of collinear field products 
by a subscript ``$s_i$''. It is also convenient to 
write down the operator basis independent of the 
Dirac structure $\Gamma_i$ of the QCD current. In this 
case we assume that the short-distance coefficients 
are Lorentz-tensors $C_{ij\ldots}$, where the dots 
stand for further indices
on the effective current. We can always decompose these into
scalar functions using $\nm^\mu,\,v^\mu,
\,g^{\mu\nu},$ and $i\eps^{\mu\nu\rho\sigma}$. The scalar functions 
depend of course on the Dirac structure of the QCD current.

\paragraph{\it Leading order.}
At leading order the only possible SCET currents are
\be
J^{(0)}=(\bar\xi W_c)_s\Gamma_j h_v.
\ee
Here $\Gamma_j=\{1,\gamma_5,\gamma_{\alpha_\perp}\}$ denotes a 
basis of the four independent Dirac matrices between 
$\bar\xi$ and $h_v$. 
For a given QCD current, we would proceed to decompose the 
tensor coefficient functions into scalar ones. 
For instance, for the V-A current relevant to semi-leptonic decay 
the three tensor coefficient functions $C_{\mu}, C_{5\mu}$, and 
$C_{\mu\alpha_\perp}$ multiplying the $J^{(0)}_j$ 
may be expressed in terms of three scalar 
coefficient functions multiplying the current operators 
\begin{equation}
(\bar\xi W_c)_s (1+\gamma_5)  \Big\{\gamma^\mu,v^\mu,
\frac{n_-^\mu}{n_- v}\Big\} h_v.
\end{equation}

\paragraph{\it Next-to-leading order.}
At relative order $\lambda$ the basis consists of the 
currents
\begin{eqnarray}
J^{(1)}_1 &=&(\bar\xi W_c)_s\Gamma_j (x_\perp D_s h_v)
\nonumber\\
J^{(1)}_2 &=&(\bar\xi W_c)_s 
\,i\overleftarrow{\partial}_{\perp}^\mu \Gamma_j h_v
\nonumber\\
J^{(1)}_3 &=&(\bar\xi W_c)_{s_1} [W_c^\dag i D_{\perp c}^\mu W_c]_{s_2}
\Gamma_j h_v.
\end{eqnarray} 
The above currents should be multiplied by an overall factor $1/m_b$
to make the coefficient functions dimensionless. The basis of operators 
with scalar coefficient functions can be found in 
\cite{Pirjol:2002km, Beneke:2004rc,Hill:2004if}. Here we 
use a basis where the 
transverse derivative is taken outside the collinear Wilson line. 
This has the advantage that only the third operator has a 
tree level matrix element with a transverse collinear gluon.

The first two operators are ``two-body'', that is, they 
depend only on $x_-$ and one other position $x+s n_+$. The 
coefficient functions of any two-body operator to any order in the 
$\lambda$ expansion can be related to the coefficient function 
of the leading-power currents. The reason for this is that 
they all follow from matching  the QCD matrix 
element $\langle q|J_i| b\rangle$ to SCET, and the number of 
independent form factors needed to parameterize the QCD matrix 
element is equal to the number of independent currents 
at leading power \cite{Beneke:2004rc}. The coefficients of the 
two-body operators follow from the expansion of the full QCD 
form factors.
The operator $J^{(1)}_3$ is a three-body operator, whose coefficient 
is determined by the expansion of $\langle q g|J_i| b\rangle$ with a 
transverse gluon to leading power. 

\paragraph{\it Next-to-next-to-leading order.}
Many structures are possible at this order.  We find 
\begin{eqnarray}
J^{(2)}_{1} &=&\frac{1}{2} \,(\bar\xi W_c)_s\Gamma_j 
\big(n_- x n_+ D_s h_v\big)\nonumber\\
J^{(2)}_{2} &=&\frac{1}{2} \,(\bar\xi W_c)_s\Gamma_j 
\big(x_{\perp\mu}  x_{\perp\nu} D_s^\mu D_s^\nu  h_v\big)
\nonumber\\
J^{(2)}_{3} &=&(\bar\xi  W_c)_s\,i\overleftarrow{\partial}_{\perp}^\mu\Gamma_j 
(x_\perp D_s h_v)
\nonumber\\
J^{(2)}_4 &=&(\bar\xi W_c)_s\Gamma_j (i D_s^\mu h_v) 
\nonumber\\
J^{(2)}_5 &=&(\bar\xi W_c)_s \,i n_- \overleftarrow{D}_s \Gamma_j h_v
\nonumber\\
J^{(2)}_6 &=&(\bar\xi W_c)_s \,i\overleftarrow{\partial}_{\perp}^\mu 
i\overleftarrow{\partial}_{\perp}^\nu \Gamma_j h_v
\nonumber\\
J^{(2)}_{7} &=&(\bar\xi W_c)_{s_1} [W_c^\dag i D_{\perp c}^\mu W_c]_{s_2}
\Gamma_j (x_\perp D_s h_v) \nonumber\\
J^{(2)}_8 &=&(\bar\xi W_c)_{s_1} \,i\overleftarrow{\partial}_{\perp}^\mu
[W_c^\dag i D_{\perp c}^\nu W_c]_{s_2}
\Gamma_j h_v \nonumber\\
J^{(2)}_9 &=&(\bar\xi W_c)_{s_1} \Big[i \,\partial_{\perp}^\mu 
(W_c^\dag i D_{\perp c}^\nu W_c)_{s_2}\Big] \Gamma_j h_v
\nonumber\\
J^{(2)}_{10} &=&(\bar\xi W_c)_{s_1} [W_c^\dag i n_- D  W_c]_{s_2}
\Gamma_j h_v
\nonumber\\
J^{(2)}_{11} &=&
(\bar\xi W_c)_{s_1} \Big[[W_c^\dag i D_{\perp c}^\mu W_c]_{s_2},
 [W_c^\dag i D_{\perp c}^\nu W_c]_{s_3}\Big] \Gamma_j h_v
\nonumber\\
J^{(2)}_{12} 
&=&(\bar\xi W_c)_{s_1} \Big\{[W_c^\dag i D_{\perp c}^\mu W_c]_{s_2},
 [W_c^\dag i D_{\perp c}^\nu W_c]_{s_3}\Big\} \Gamma_j h_v
\nonumber\\
J^{(2)}_{13} &=&(\bar\xi W_c)_{s_1} \mbox{tr}\Big[
[W_c^\dag i D_{\perp c}^\mu W_c]_{s_2}
 [W_c^\dag i D_{\perp c}^\nu W_c]_{s_3}\Big] \Gamma_j h_v
\nonumber\\
J^{(2)}_{14} &=&\Big[(\bar\xi W_c)_{s_1}\Gamma_j h_v\Big]\,
\Big[(\bar\xi W_c)_{s_2} \frac{\slash{n}_+}{2}\,\Gamma_{j^\prime} 
(W^\dagger_c\xi)_{s_3}\Big]
\nonumber\\
J^{(2)}_{15} &=&\Big[(\bar\xi W_c)_{s_1}\Gamma_j T^A h_v\Big]\,
\Big[(\bar\xi W_c)_{s_2} \frac{\slash{n}_+}{2}\,\Gamma_{j^\prime} 
T^A (W^\dagger_c\xi)_{s_3}\Big].
\end{eqnarray}
The trace refers to colour, and an 
overall factor $1/m_b^2$ should be included to make the 
coefficient functions dimensionless.  
The argument for the completeness of this basis is as follows. 
The QCD matrix element $\langle q|J_i| b\rangle$ is expanded 
to order $\lambda^2$, which involves the transverse momentum squared,  
$n_-$ times the collinear momentum, or the heavy quark residual momentum. 
Together with the terms from the multipole expansion of the heavy quark 
field, this accounts 
for the two-body operators $J^{(2)}_{1-6}$. 
Next, we can have $\langle q g|J_i| b\rangle$ 
expanded to first order in the transverse 
momentum of the collinear quark or the 
collinear gluon, which gives $J^{(2)}_{7-9}$, with $J^{(2)}_{7}$ 
coming from the multipole expansion. 
The coefficient of  $J^{(2)}_{7}$ is related to that of the 
order $\lambda$ three-body operator $J_3^{(1)}$.  Some of the scalar 
coefficients of  $J^{(2)}_{8,9}$ are also related to $J_3^{(1)}$. 
However, contrary to the case of two-body operators it is no 
longer true that the coefficients of all sub-leading three-body 
operators are related to the leading one, because the number 
of invariant form factors needed to parameterize  
$\langle q g|J_i| b\rangle$ in QCD is larger than the number of the 
leading  three-body currents. We also get new three-body operators from 
$\langle q g|J_i| b\rangle$ with a gluon corresponding 
to a $\nm A_c$ field or $\nm A_s$, but gauge 
invariance requires that the soft gluon is part of 
a covariant derivative, so we get the three-body operator
$J^{(2)}_{10}$. In fact, $J_5^{(2)}$ and $J_{10}^{(2)}$ could be 
eliminated using the collinear quark and gluon equations of motion. 
Finally, we get four-body 
operators from $\langle q g g|J_i| b\rangle$ with two transverse 
collinear gluons 
or from $\langle q q\bar q|J_i| b\rangle$ with three collinear 
(anti-)quarks. There are several possible colour structures, 
which can be chosen as in $J^{(2)}_{11-15}$. 

\paragraph{\it {}}It is useful to gain some intuition as to how
this general set of heavy-light currents translates
into terms in the factorization formula.  First, it
is obvious that insertions of heavy-light currents
which contain no soft gluon fields simply build up 
a set of power-suppressed jet-functions convoluted
with the leading order shape-function.  This includes
two insertions of $J_{2,3}^{(1)}$ or single insertions
of $J_{6, 8-15}^{(2)}$.  If we work at tree level, 
only two-body operators can contribute to the current 
correlator, and we need only 
consider two insertions of $J_{1,2}^{(1)}$ or a single
insertion of $J_{1-6}^{(2)}$. The full set of three- and four-body 
operators is only needed when one aims at an accuracy 
$\alpha_s\LQCD/m_b$ in the calculation of the hadronic 
tensor.

\section{Basis of shape-functions}
\label{sec:basis}

In this section we will collect the results for   
the independent matrix elements of soft fields (``shape-functions'')  
needed to parameterize the hadronic tensor at order 
$\lambda^2$. The possible time-ordered products  
that build up the current correlator to this order 
are   
\begin{eqnarray}
\label{eq:combos}
a) && J^{(0)} J^{(2)}_k + \,\mbox{sym.}
\nonumber\\
b) && J^{(1)}_k J^{(1)}_l 
\nonumber\\
c) && J^{(0)} J^{(1)}_k {\cal L}^{(1)} + \,\mbox{sym.}
\nonumber\\
d) && J^{(0)} J^{(0)} {\cal L}^{(2)} 
\nonumber\\
e) && J^{(0)} J^{(0)} {\cal L}^{(1)}  {\cal L}^{(1)}.
\end{eqnarray}
Inspection of 
the effective currents and Lagrangian shows that 
from these products we obtain the following soft
operators (leaving out colour and spinor indices): 
\begin{enumerate} 
\item From a) and b)
\begin{eqnarray}
&&(\bar h_v Y) (Y^\dagger h_v),\, 
(\bar h_v  Y) (Y^\dagger i D_s^\mu h_v),\,
(\bar h_v  (-i)\overleftarrow{D}_s^\mu Y) (Y^\dagger h_v),\,
\nonumber\\
&&(\bar h_v (-i)\overleftarrow{D}_s^{\mu_\perp}
 (-i)\overleftarrow{D}_s^{\nu_\perp} Y) (Y^\dagger  h_v). 
\label{bilocal}
\end{eqnarray}
We leave out operators related by hermitian conjugation. 
The convention is such that fields in different parentheses  
stand at different positions $0, x_-, z_-,\ldots$ and that colour indices 
in brackets are contracted. Hence all these 
terms are bi-local. 

\item From the single Lagrangian insertions c) and d) in addition
\begin{eqnarray}
&&(\bar h_v Y) (Y^\dagger  i g n_-^\nu F^s_{\mu_\perp\nu}Y) 
(Y^\dagger  h_v), \,
(\bar h_v  (-i) \overleftarrow{D}_s^{\rho_\perp} Y) (Y^\dagger i g n_-^\nu 
F^s_{\mu_\perp\nu}Y) (Y^\dagger  h_v),
\nonumber\\
&&(\bar h_v Y) (Y^\dagger  i g n_+^\mu n_-^\nu F^s_{\mu\nu}Y) 
(Y^\dagger  h_v), \,
(\bar h_v Y) (Y^\dagger  i g F^s_{\mu_\perp\nu_\perp}Y) 
(Y^\dagger  h_v), \,
\nonumber\\
&&(\bar h_v Y) (Y^\dagger  [i D_s^{\rho_\perp}, 
i g n_-^\nu F^s_{\mu_\perp\nu}] Y) (Y^\dagger  h_v), \,
(\bar h_v Y) (Y^\dagger  [i n_- D_s,i g F^s_{\mu_\perp\nu_\perp}] Y) 
(Y^\dagger  h_v), \,
\nonumber\\
&& (\bar h_v Y) (Y^\dagger h_v) \,i\int d^4 z \,{\cal L}^{(2)}_{\rm HQET}(z). 
\label{trilocal}
\end{eqnarray}
These are tri-local (with exception of the last line). The
second-to-last operator comes only from the Yang-Mills Lagrangian 
${\cal L}_{\rm YM}^{(2)}$.
\item From the double insertions e) 
\begin{eqnarray}
&&(\bar h_v Y) (Y^\dagger i g n_-^\nu F^s_{\mu_\perp\nu} Y) 
(Y^\dagger i g n_-^\sigma F^s_{\rho_\perp\sigma} Y) (Y^\dagger  h_v), 
\nonumber\\
&&(\bar h_v Y) (Y^\dagger  h_v)  (\bar q_s Y) (Y^\dagger  q_s). 
\label{tetralocal}
\end{eqnarray}
These are tetra-local. 
\end{enumerate}
We note immediately the increase in complexity of the 
matrix elements needed to parameterize the $1/m_b$ corrections 
to all orders in perturbation theory, since there appear tetra-local 
light-cone operators including a four-quark operator  
which has not yet been discussed in the literature.  

It is convenient to parameterize the 
shape-functions in a covariant derivative basis. Reinstating 
colour and spinor indices, the field 
products listed above can be derived from 
\bea
&&(\bar h_v Y)(x_-)_{a\alpha} (Y^\dagger h_v)(0)_{b\beta},
\nonumber \\
&& (\bar h_v Y)(x_-)_{a\alpha} (Y^\dagger iD_s^\mu Y)(z_-)_{cd} 
(Y^\dagger h_v)(0)_{b\beta},
\nonumber \\
&& (\bar h_v Y)(x_-)_{a\alpha} (Y^\dagger iD_{s\perp}^\mu Y)(z_{1-})_{cd}
(Y^\dagger iD_{s\perp}^\nu Y)(z_{2-})_{ef} (Y^\dagger
h_v)(0)_{b\beta}, 
\label{derivativebasis}
\eea
and the expressions in the last lines of (\ref{trilocal},\ref{tetralocal}), 
respectively, using in particular the identities
$Y^\dagger  i n_- D_s Y = i n_-\partial$, 
\begin{equation}
Y^\dagger  i g n_-^\nu F^s_{\mu\nu}Y = 
-\left[i n_-\partial, Y^\dagger iD^s_\mu Y \right] = 
-\left[i n_-\partial Y^\dagger [iD^s_\mu Y] \right].
\label{fieldtocov}
\end{equation}
The derivatives are meant to act 
on everything to the right irrespective of position
argument, but derivatives in square brackets (not to be confused 
with commutators) as in the last expression act
only within the square brackets. With 
this convention 
\begin{equation}
(Y^\dagger iD_s^\mu Y)(z)_{cd} = i\partial^\mu \delta_{cd} + 
(Y^\dagger [iD_s^\mu Y])(z)_{cd}. 
\label{covD}
\end{equation}
Note that since in the second term the derivative acts only on $Y$, 
this term is a colour octet (in light-cone gauge, $n_- A_s=0$, and we have 
$Y^\dagger [iD_s^\mu Y] = g A_s^\mu$). The first term is a colour-singlet, 
but does not depend on $z$. Hence if, for instance, 
$(Y^\dagger iD_s^\mu Y)(z)_{cd}$ appears in a tri-local matrix element, 
the colour-singlet part is only bi-local. 

We decompose the matrix elements of (\ref{derivativebasis}) into a 
set of scalar shape-functions. The only possible Dirac structures between 
two static quark fields $\bar h_v \ldots h_v$ are 
$1$ and $(\gamma^\mu-\slash{v} v^\mu)\gamma_5$. Defining  
$\epsilon^\perp_{\mu\nu}\equiv i \epsilon_{\mu
\nu\rho\sigma} n_-^\rho v^\sigma$, we write  
(recall that $x_-^\mu=x_+ n_-^\mu$)
\begin{eqnarray}
&&\langle \bar B|(\bar h_v Y)(x_-)_{a\alpha} (Y^\dagger h_v)(0)_{b\beta}|
\bar B\rangle = 
 \frac{\delta_{ba}}{N_c}\,
 \frac{1}{2}\left(\frac{1+\slash{v}}{2}\right)_{\!\beta\alpha}
 \tilde{S}(x_+),
 \\[0.3cm]
&&\langle \bar B|(\bar h_v Y)(x_-)_{a\alpha} (Y^\dagger h_v)(0)_{b\beta}
 \,i \int d^4 z \,{\cal L}^{(2)}_{\rm HQET}(z) |\bar B\rangle = 
 \nonumber\\&&\hspace*{2cm}
 \frac{1}{2 m_b}\,\frac{\delta_{ba}}{N_c}\,
 \frac{1}{2}\left(\frac{1+\slash{v}}{2}\right)_{\!\beta\alpha}
 \left[\tilde{s}_{\rm kin}(x_+)+C_{\rm mag}(m_b/\mu) 
       \tilde{s}_{\rm mag}(x_+)\right],
 \\[0.3cm]
&&\langle \bar B|(\bar h_v Y)(x_-)_{a\alpha} 
 (Y^\dagger i D_s^{\mu} Y)(z_-)_{cd} 
 (Y^\dagger h_v)(0)_{b\beta}|\bar B\rangle = 
 \nonumber\\&&\hspace*{2cm}
 \frac{1}{2}\left(\frac{1+\slash{v}}{2}\right)_{\!\beta\alpha}
 \Bigg\{
 \frac{\delta_{ba} \delta_{cd}}{N_c}
 \left[-i\tilde S^\prime(x_+)v^\mu + \Big(i\tilde S^\prime(x_+)-
       \tilde T_1(x_+,0)\Big) n_-^\mu\right] 
 \nonumber\\&&\hspace*{4cm}
 + \, \frac{2 \,T^A_{ba} T^A_{cd}}{N_c^2-1}
 \left[\tilde T_1 (x_+,z_+) n_-^\mu\right]
 \bigg\}
\nonumber\\&& \hspace*{2cm}
 + \, \frac{1}{2}\left(\frac{1+\slash{v}}{2}\gamma_{\rho_\perp}
 \gamma_5\frac{1+\slash{v}}{2}\right)_{\!\beta\alpha}
 \frac{\epsilon_\perp^{\mu\rho}}{2}\,
 \bigg\{
 \frac{\delta_{ba} \delta_{cd}}{N_c} 
       \Big(\tilde t(x_+)-\tilde T_2(x_+,0)\Big)  
 \nonumber\\&& \hspace*{4cm}
 + \,\frac{2 \,T^A_{ba} T^A_{cd}}{N_c^2-1}\,
 \tilde{T}_2(x_+,z_+) 
 \Bigg\},
 \label{oneDshape} \\[0.3cm]
&&\langle \bar B|(\bar h_v Y)(x_-)_{a\alpha} 
 (Y^\dagger iD^s_{\mu_\perp} Y)(z_{1-})_{cd} 
 (Y^\dagger i D^s_{\nu_\perp} Y)(z_{2-})_{ef} 
 (Y^\dagger h_v)(0)_{b\beta}|\bar B\rangle = 
 \nonumber\\&&\hspace*{2cm}
 \frac{1}{2}\left(\frac{1+\slash{v}}{2}\right)_{\!\beta\alpha}
 \frac{g^\perp_{\mu\nu}}{2}\,
 \Bigg\{ \frac{\delta_{ba}\delta_{cd}\delta_{ef}}{N_c}\,
   \tilde u_1(x_+) 
 \nonumber\\&&\hspace*{2cm}+\, 
 \frac{2 \,\delta_{cd} T^A_{ba} T^A_{ef}}{N_c^2-1} \,
   \frac{1}{2}\left[\tilde U_1(x_+, z_{2+})-\tilde U_2(x_+,z_{2+})-
   \tilde{{\cal U}}_1(x_+,z_{2+},z_{2+})\right]
 \nonumber\\&&\hspace*{2cm}+\, 
 \frac{2 \,\delta_{ef} T^A_{ba} T^A_{cd}}{N_c^2-1} \,
   \frac{1}{2}\left[\tilde U_1(x_+, z_{1+})+\tilde U_2(x_+,z_{1+})-
   \tilde{{\cal U}}_1(x_+,z_{1+},z_{1+})\right]
 \nonumber\\&&\hspace*{2cm}+\, 
 \frac{2 \,\delta_{ba} T^A_{cd} T^A_{ef}}{N_c^2-1}\,
 \left[\tilde{{\cal U}}_1(x_+,z_{1+},z_{2+})-
       \tilde{{\cal U}}_2(x_+,z_{1+},z_{2+})-
       \tilde{{\cal U}}_3(x_+,z_{1+},z_{2+})\right] 
 \nonumber\\&&\hspace*{2cm}-\, 
 \frac{4i f^{ABC} T^A_{cd} T^B_{ef}T^C_{ba}}{N_c(N_c^2-1)}\,
  \tilde{{\cal U}}_2(x_+,z_{1+},z_{2+}) + 
 \frac{4 N_c d^{ABC} T^A_{cd} T^B_{ef}T^C_{ba}}{(N_c^2-1)(N_c^2-4)}\,
 \tilde{{\cal U}}_3(x_+,z_{1+},z_{2+})
 \Bigg\}
 \nonumber\\&&\hspace*{2cm}
 +\,\left(-\frac{1}{2}\right)\left(\frac{1+\slash{v}}{2}\slash{\,n}_-
 \gamma_5\frac{1+\slash{v}}{2}\right)_{\!\beta\alpha}
 \frac{\epsilon^\perp_{\mu\nu}}{2}\, 
 \Bigg\{ \tilde u_1\to 0,\,\, \tilde U_{1,2}\to \tilde U_{3,4},\,\,
 \nonumber\\&&\hspace*{4cm} 
 \tilde {\cal U}_{1,2,3}\to  \tilde {\cal U}_{4,5,6} \Bigg\},
 \label{twoDshape}\\[0.3cm]
&&\langle \bar B|(\bar h_v Y)(x_-)_{a\alpha} 
 (Y^\dagger h_v)(0)_{b\beta}\,(\bar q_s Y)(z_{1-})_{c\gamma} 
 (Y^\dagger q_s)(z_{2-})_{d\delta}|\bar B\rangle = 
 \nonumber\\&&\hspace*{2cm}
 \frac{\delta_{da}\delta_{bc}}{N_c^2}\Bigg\{
 \frac{1}{2}\left(\frac{1+\slash{v}}{2}\right)_{\!\beta\alpha} \Big[
 \delta_{\delta\gamma}\,\tilde{{\cal V}}_1(x_+,z_{1+},z_{2+})+
 \slash{v}_{\delta\gamma}\,\tilde{{\cal V}}_2(x_+,z_{1+},z_{2+})
 \nonumber\\&&\hspace*{4cm}+\,
 {\slash{n}_-}_{\delta\gamma}\,\tilde{{\cal V}}_3(x_+,z_{1+},z_{2+})+
 (\slash{n}_-\slash{v})_{\delta\gamma}\,
    \tilde{{\cal V}}_4(x_+,z_{1+},z_{2+})\Big] 
 \nonumber\\&&\hspace*{3cm}+\,
 \frac{1}{2}\left(\frac{1+\slash{v}}{2}\gamma^{\mu}
 \gamma_5\frac{1+\slash{v}}{2}\right)_{\!\beta\alpha} \Big[
 (\gamma_\mu\gamma_5)_{\delta\gamma}\,\tilde{{\cal V}}_5(x_+,z_{1+},z_{2+}) 
 \nonumber\\&&\hspace*{4cm}+\,
 (\slash{v}\gamma_\mu\gamma_5)_{\delta\gamma}\,
     \tilde{{\cal V}}_6(x_+,z_{1+},z_{2+})+
 (\slash{n}_-\gamma_\mu\gamma_5)_{\delta\gamma} \,
     \tilde{{\cal V}}_7(x_+,z_{1+},z_{2+})
 \nonumber\\&&\hspace*{4cm}+\,
 (\slash{n}_-\slash{v}\gamma_\mu\gamma_5)_{\delta\gamma}\,
 \tilde{{\cal V}}_8(x_+,z_{1+},z_{2+})\Big] \Bigg\} 
 \nonumber\\&&\hspace*{2cm}+\,
 \frac{4\,T^A_{da} T^A_{bc}}{(N_c^2-1)}\Bigg\{
 \frac{1}{2}\left(\frac{1+\slash{v}}{2}\right)_{\!\beta\alpha} \Big[
 \tilde{{\cal V}}_{1-4}\to \tilde{{\cal V}}_{9-12}\Big] 
 \nonumber\\&&\hspace*{3cm}+\,
 \frac{1}{2}\left(\frac{1+\slash{v}}{2}\gamma^{\mu}
 \gamma_5\frac{1+\slash{v}}{2}\right)_{\!\beta\alpha} \Big[
 \tilde{{\cal V}}_{5-8}\to \tilde{{\cal V}}_{13-16}\Big] \Bigg\}.  
\label{fourquarkshape}
\end{eqnarray}
This decomposition makes no assumption on whether the external state 
is a $\bar B$ meson, a $\bar B^*$ meson or a $b$-hadron, but  
it assumes that it is averaged over polarizations, so that 
the only available vectors are $v$ and $n_-$.  Our notation 
uses lowercase letters for shape-functions
depending on a single variable, capital letters for those depending 
on two variables, and calligraphic letters for those depending on 
three variables.  The only exception is our notation for the leading 
order shape-function ${\tilde S}(x_+)$, where we use a capital letter 
even though it depends only on a single variable. The multi-locality
of a given shape-function is determined by decomposing the
colour structure of the covariant derivatives  
$(Y^\dagger iD_s^\mu Y)(z)_{cd}$ as  in (\ref{covD}),
and then using that the singlet component has 
a lower degree of non-locality than the octet component.  
The parameterization is chosen such that the colour 
contractions with the tree-level jet-functions,  
as well as the limits  $z=0$ in (\ref{oneDshape})
and $z_1=z_2$ in (\ref{twoDshape}), take a simple form. For instance,
writing the combination $\tilde t(x_+)-\tilde T_2(x_+,0)$ for the
colour singlet term in the fourth line of 
(\ref{oneDshape}) ensures that only
$\tilde t(x_+)$ appears at tree-level.
We used the heavy quark equation of motion to reduce the number of 
independent functions. The gluon field equation can be used 
to eliminate $(Y^\dagger i n_+D_s Y)(z)_{cd}$. This does not 
lead to a simplification in practice, and we have therefore 
kept $\tilde T_1(x_+, z_+)$ as a basis shape-function. 
$\tilde S^\prime$ is the derivative of $\tilde S$ with 
respect to $x_+$.

It follows that to order $1/m_b$, but 
to arbitrary order in $\alpha_s$ in the coefficient functions, 
the differential decay rates depend on a large number of 
multi-local shape-functions. Fortunately, the structure of 
the result is much simpler in the tree approximation as we 
shall see below. In addition to displaying the general set 
of shape-functions, the above enumeration of soft matrix elements 
allows us to clarify in more detail 
the structure of the convolutions in the factorization formula 
(\ref{eq:masfact}). Since the effective currents contain at most 
three products of collinear fields at different positions, 
the number of collinear convolutions is at most two.   
The shape-functions are at most tetra-local, so the maximal number 
of soft convolution integrals at order $1/m_b$ is three. This results 
in the structure 
\begin{eqnarray}
 T &=& H \cdot \,\J(\omega)\otimes S(\omega)
\nonumber \\
&& +\, \sum H(u_1,u_2)\otimes 
\J(u_1,u_2;\omega)\otimes S(\omega)
+ \sum H(u)\otimes 
\J(u;\omega_1, \omega_2)\otimes S(\omega_1,\omega_2)
\nonumber \\
&& +\, \sum H \cdot \, 
\J(\omega_1, \omega_2,\omega_3)\otimes S(\omega_1,\omega_2,\omega_3) 
+\ldots 
\label{eq:masfact2},
\end{eqnarray}
where the ellipses denote $1/m_b^2$ terms not considered here, 
and for each term the most complicated structure is shown. The 
momentum space shape-functions are defined in terms of the 
coordinate space shape-functions given above by the Fourier transform 
(\ref{defshape}). 
The variable $\omega_i$ corresponds to $n_- k_i$, where $k_i$ is the 
(outgoing) momentum of soft fields.

\section{Example of factorization at order $1/m_b$}
\label{sec:example}

We now have all the ingredients needed to calculate the factorization
formula to order $\lambda^2\sim 1/m_b$.  To illustrate the step-by-step
procedure outlined in Section \ref{sec:fact}, we  work out 
the contribution from the time-ordered product 
\begin{equation}
T\Big\{J^{(0)\dagger}(x) J_{2}^{(1)}(0) \,i\!\int d^4 z \, {\cal L}^{(1)}_\xi(z)
\Big\}
\label{exampleproduct}
\end{equation}
as an example. All other terms can be treated in a similar
way. However, there is a large number of them, and we do not list them
in this paper explicitly. 

Specifically, we consider the currents 
\begin{equation}
J^{(0)} = (\bar\xi W_c)_{s_1} \gamma^\mu h_v,\qquad
J_2^{(1)} = (\bar\xi W_c)_{s_1} i\overleftarrow{\slash{\partial}}_{\!\!\perp}
\gamma^\mu h_v,
\end{equation}
which appear in the SCET expansion of the vector current. 
In the first step we obtain the corresponding contribution to the 
current correlation function $T^{\mu\nu}$ ($p=m_b v-q$),
\bea
T^{\mu\nu}&=& i\int d^4x\, e^{ip\cdot x}\int d{\hat s}_1
d{\hat s}_2 \,\tilde{C}^{(0)\star}({\hat s}_1)\tilde{C}^{(1)}_2({\hat s}_2)
\nonumber\\
&& 
T\Big\{\bar h_v(x_-) \gamma^{\mu} (W_c^\dagger \xi)(x+s_1\np)
\,\Big[(\bar\xi W_c)(s_2\np)i\overleftarrow{\slash{\partial}}_{\!\!\perp}
\Big]\gamma^\nu\,h_v(0)
\nonumber\\ 
&&\hspace{1cm}
i \int d^4 z\,
(\bar{\xi} W_c)(z) z_{\perp}^\lambda n_-^\rho g F^s_{\lambda\rho}(z_-)
\frac{\slash{n}_+}{2} (W_c^\dagger\xi)(z)\Big\}.
\eea
The coefficient function $\tilde{C}^{(1)}_2$ can be related 
to $\tilde{C}^{(0)}$ as explained in Section~\ref{sec:currents}, 
but the detailed form of this relation is not important for the
following. We now use translation invariance to shift all fields 
by the amount $-s_2 n_+$, then perform the change of variables 
$x\to x+(s_2-s_1) n_+$, $z\to z+s_2 n_+$. This does not affect the 
position of the soft fields, since $[s_i n_+]_-=0$. The integrations 
over $\hat{s}_{1,2}$ can then be performed to obtain 
\bea
T^{\mu\nu}&=& H(n_+ p/m_b) \, i\int d^4x\, e^{ip\cdot x}
\,T\Big\{\bar h_v(x_-) \gamma^{\mu} (W_c^\dagger \xi)(x)
\,\Big[(\bar\xi W_c)(0)i\overleftarrow{\slash{\partial}}_{\!\!\perp}
\Big]\gamma^\nu\,h_v(0)
\nonumber\\ 
&&\hspace{1cm}
i \int d^4 z\,
(\bar{\xi} W_c)(z) z_{\perp}^\lambda n_-^\rho g F^s_{\lambda\rho}(z_-)
\frac{\slash{n}_+}{2} (W_c^\dagger\xi)(z)\Big\}
\eea
with $H(n_+ p/m_b) = C^{(0)}(n_+ p/m_b)C_2^{(1)}(n_+ p/m_b)$. 
($C^{(0)}$ is real.) The 
effects of the hard scale have turned into a multiplicative 
factor $H$ rather than a convolution, because both effective 
currents were only two-body. Next we perform the collinear field 
redefinition that decouples collinear and soft fields. We also use  
(\ref{fieldtocov}) and integrate by parts to 
let the $i\nm\partial_{(z)}$ act on the collinear fields. 
The result is 
\bea
\langle \bar B|T^{\mu\nu}|\bar B\rangle &=& 
H(n_+ p/m_b) \, i\int d^4x d^4z \, e^{ip\cdot x} \,z^{\rho_\perp} \,
\nonumber\\
&&\hspace{0cm}
\langle \bar B|(\bar h_v Y)(x_-)_{a\alpha} 
 (Y^\dagger [i D^s_{\rho_\perp} Y])(z_-)_{cd} 
 (Y^\dagger h_v)(0)_{b\beta}|\bar B\rangle
\label{eq112}\\ 
&&\hspace{0cm}
i\nm\partial_{(z)}
\langle 0|\,
T\Big\{(\gamma^{\mu} W_c^\dagger \xi)(x)_{a\alpha}
\,\Big[(\bar\xi W_c)i\overleftarrow{\slash{\partial}}_{\!\!\perp}
\gamma^\nu\Big](0)_{b\beta}
(\bar{\xi} W_c)(z)_c \frac{\slash{n}_+}{2} (W_c^\dagger\xi)(z)_d\Big\}
|0\rangle. 
\nonumber
\eea
We now insert (\ref{oneDshape}) for the $\bar B$ meson matrix element 
of the soft fields. Only the colour component $T^A_{ba} T^A_{cd}$ 
contributes, and 
we are left with the term parameterized by the
tri-local shape-function $\tilde T_2$. Going to momentum space 
shape-functions, we find 
\bea
\langle \bar B|T^{\mu\nu}|\bar B\rangle &=& 
\frac{2}{N_c^2-1}\,\frac{\epsilon^\perp_{\rho\sigma}}{2}
\,\frac{1}{2}\left(\gamma^\nu\frac{1+\slash{v}}{2}\gamma^{\sigma_\perp}
\gamma_5\gamma^\mu\right)_{\beta\alpha}\,H(n_+ p/m_b) \,
\nonumber\\
&&\hspace{0cm}
\int d\omega_1 d\omega_2\,T_2(\omega_1,\omega_2)
\int d^4x d^4z \, e^{ip\cdot x} \,e^{-i\omega_1 x_+}\,
e^{-i\omega_2 z_+}\,(-i) z^{\rho_\perp} \,
\label{eq113}
\\ 
&&\hspace{0cm}
i\nm\partial_{(z)}
\langle 0|\,
T\Big\{
\Big[(\bar\xi W_c)i\overleftarrow{\slash{\partial}}_{\!\!\perp}
\Big](0)_{\beta}T^A(W_c^\dagger \xi)(x)_{\alpha}
(\bar{\xi} W_c  \frac{\slash{n}_+}{2}T^A W_c^\dagger\xi)(z)\Big\}
|0\rangle.
\nonumber
\eea
The integral over $x$ and $z$ defines the momentum space jet
function $\J$. In the frame with $p_\perp=0$ it can depend on 
the external momenta $n_+ p$, $p^2$, and the convolution 
variables $\omega_{1,2}$,
and the integral equals 
\begin{equation}
\frac{N_c^2-1}{2} \left(\frac{\slash{n}_-}{2}\gamma^{\rho_\perp}
\!\right)_{\alpha\beta} \J(n_+ p, p^2;\omega_1,\omega_2).
\end{equation}
Inserting this into (\ref{eq113}) we obtain the final result 
\begin{eqnarray}
\langle \bar B|T^{\mu\nu}|\bar B\rangle &=& 
\frac{\epsilon^\perp_{\rho\sigma}}{2}
\,\frac{1}{2}\mbox{tr}\left(\frac{\slash{n}_-}{2}\gamma^{\rho_\perp}
\gamma^\nu\frac{1+\slash{v}}{2}\gamma^{\sigma_\perp}
\gamma_5\gamma^\mu\right)
\nonumber\\
&&\hspace{0cm}
H(n_+ p/m_b) \,\int d\omega_1 d\omega_2\,
T_2(\omega_1,\omega_2)
 \,\J(n_+ p, p^2;\omega_1,\omega_2),
\end{eqnarray}
which expresses the contribution from (\ref{exampleproduct}) 
to the current correlator as a product of a hard coefficient 
and a two-fold convolution of a shape-function and a jet-function. 

It is instructive to examine the result in the tree approximation 
for the jet-function.  In this approximation 
\be
\J^{{\rm tree}}(n_+ p, p^2;\omega_1,\omega_2)=
\omega_2\,\frac{i n_+ p}{p_{\omega_1}^2}
\frac{i n_+ p}{p_{\omega_{12}}^2} = 
\frac{n_+ p}{p_{\omega_1}^2} - 
\frac{n_+ p}{p_{\omega_{12}}^2},
\label{treejet}
\ee
where we used  $n_+ p \,\omega_2 = p_{\omega_1}^2-p_{\omega_{12}}^2$ 
($p_{\omega_{1\ldots n}}=p-(\omega_1 +\ldots+\omega_n) n_+/2$).
Each of the two terms after the second equality depends only on a single 
combination of the $\omega_{1,2}$, which allows us to write 
\bea
&&\int d\omega_1 d\omega_2\,
T_2(\omega_1,\omega_2)
 \,\J^{{\rm tree}}(n_+ p, p^2;\omega_1,\omega_2)
\nonumber\\
&&\hspace*{3cm} =  
\int d\omega \,\frac{n_+ p}{p_\omega^2} 
\int d\omega^\prime \Big(T_2(\omega-\omega^\prime,\omega^\prime)-
T_2(\omega,\omega^\prime)\Big).
\eea
The structure of this result implies in coordinate space that 
rather than the tri-local matrix element (\ref{oneDshape}) 
parameterized by a function of two variables $\tilde T_2(x_+, z_+)$, 
the tree level approximation requires only the simpler combination 
$\tilde T_2(x_+, x_+)-\tilde T_2(x_+,0)$ related to the 
{\em bi-local} matrix element
\begin{equation}
\langle \bar B|(\bar h_v (-i)\overleftarrow{D}^s_{\mu_\perp} Y)(x_-)
\gamma_{\nu_\perp}\!\gamma_5 
 (Y^\dagger h_v)(0)|\bar B\rangle - 
\langle \bar B|(\bar h_v Y)(x_-)\gamma_{\nu_\perp}\!\gamma_5 
 (Y^\dagger i D^s_{\mu_\perp} h_v)(0)|\bar B\rangle.
\end{equation}
The simplification of the tree level result is due to an equation 
of motion identity behind the cancellation of propagators in 
(\ref{treejet}). We can see this directly in coordinate space 
by using the SCET equation of motion for the free-field propagator. 
Defining the contraction
\begin{equation} 
\label{contraction} 
\bar \xi(x)_{a\alpha} \xi(y)_{b\beta}  
=i\Delta(x-y) \,\delta_{ab}
\left(\frac{\slash{n}_-}{2}\right)_{\alpha\beta},  
\end{equation} 
\unitlength1cm 
\begin{picture}(0,0)(0,0.2) 
 \put(4.9,0.8){\line(0,-1){0.3}} 
 \put(4.9,0.5){\line(1,0){1.1}} 
 \put(6.0,0.5){\line(0,1){0.3}} 
\end{picture} 
\vskip-0.1cm\noindent 
the function $\Delta(z)$ satisfies 
\begin{equation}
\label{freeeom}
i n_-\partial \Delta(z) = \delta^{(4)}(z)-\frac{(i\partial_\perp)^2}
{i n_+\partial} \Delta(z).
\end{equation}
When this is used on the product of the two collinear 
propagators in (\ref{eq112}), the second term 
on the right hand side of (\ref{freeeom}) gives zero, because we work
in the frame $p_\perp=0$, and the two delta-functions produce  
the two terms $\tilde T_2(x_+, x_+)-\tilde T_2(x_+,0)$. 
We will see more of these simplifications when we work out 
the complete result in the tree approximation in the following 
section. In Section~\ref{sec:beyondtree} we ask in more generality 
whether the appearance of tetra- and tri-local shape-functions 
in the formalism could be spurious, and whether, perhaps, the 
final result could be expressed in terms of only bi- or tri-local 
matrix elements  beyond the tree approximation.

\section{Tree approximation}
\label{sec:tree}

%%%%%%%%%%%%%%%%%%%%%%%%%%%%%%%%%%%%%%%%%%%%%%%%%%%%%%%%%%%%%%%%%%%
\begin{figure}[t]
%   \vspace{-3.5cm}
%   \epsfysize=10cm
   \epsfxsize=14cm
   \centerline{\epsffile{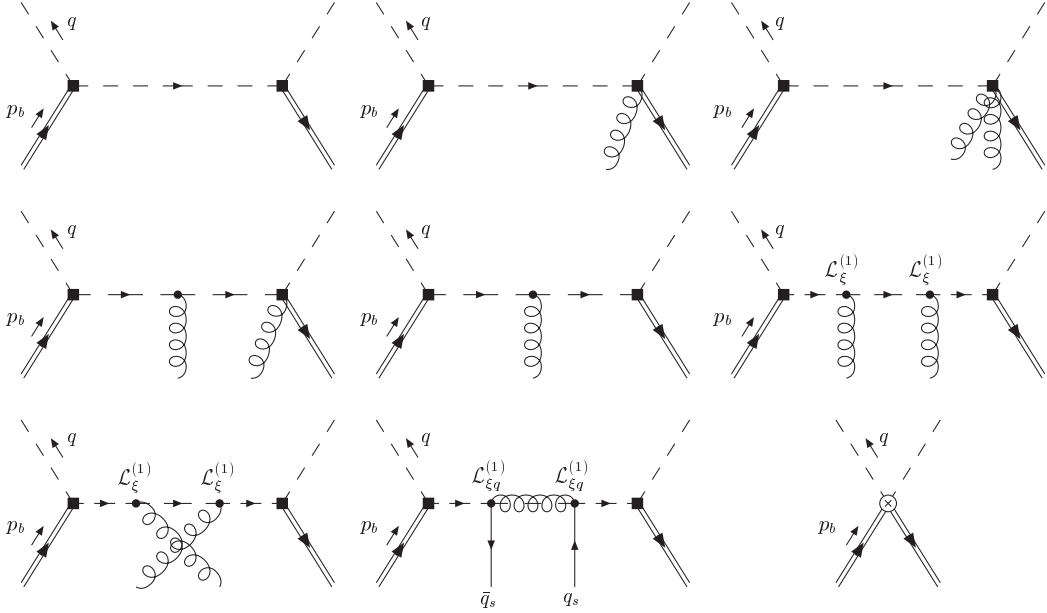}}
   \vspace*{0.5cm}
\centerline{\parbox{14cm}{\caption{\label{fig:tree}\small  
Tree diagrams contributing to the current correlator $T^{\mu\nu}$. 
Not shown are diagrams that vanish when $n_+ A_c=0$, $n_- A_s=0$, 
or are symmetric to those shown.}}}
\end{figure}
%%%%%%%%%%%%%%%%%%%%%%%%%%%%%%%%%%%%%%%%%%%%%%%%%%%%%%%%%%%%%%%%%%%

In this section we calculate the current correlator  $T^{\mu\nu}$ 
(\ref{eq:correlator}) 
and the hadronic tensor in the tree approximation including 
the $1/m_b$ power corrections. In this approximation we can 
set the collinear gluon field $A_c$ to zero except in 
${\cal L}^{(1)}_{\xi q}$, where we can draw a tree graph with 
external soft light quarks (see Figure~\ref{fig:tree}). The weak current 
$J_i = \bar{\psi} \hspace*{0.03cm}\Gamma_i  \hspace*{0.03cm}Q$ 
including the hard coefficient functions (at tree level) is 
given up to order $\lambda^2$ by \cite{Beneke:2002ph,Beneke:2002ni}
\begin{equation}
J_i = J^{(0)}+\sum_{i=1}^2 J_i^{(1)}+\sum_{i=1}^4 J_i^{(2)}
\end{equation}
with  
\begin{eqnarray}
\label{eq:currents}
J^{(0)}&=&\bar{\xi}\Gamma_i  h_v,   \nonumber\\
J^{(1)}_{1}& =& \bar{\xi}\Gamma_i x_{\perp\mu}D_{\perp s}^{\mu}h_v, 
\quad J^{(1)}_{2}=-\bar{\xi} \,i\Slash{{\ov \partial}}_{\!\perp }\frac{1}{i
  \np\ov \partial}
\frac{\slash{n}_+}{2}\Gamma_i  h_v,\nonumber\\
J^{(2)}_{1}&=&\bar{\xi}\Gamma_i \frac{\nm x}{2}\,\np D_s h_v, \quad 
J^{(2)}_{2}=\bar{\xi}\Gamma_i \frac{x_{\mu\perp}x_{\nu\perp}}{2}
D^\mu_{\perp s}\,D^\nu_{\perp s}h,\nonumber\\
J^{(2)}_{3}&=&-\bar{\xi}\,i\Slash{{\ov \partial}}_{\!\perp}
\frac{1}{i {\np\ov\partial}}
\frac{\slash{n}_+}{2}\Gamma_i x_{\perp\mu}D^\mu_{\perp s}h_v, \quad 
J^{(2)}_{4}=\bar{\xi}\Gamma_i \,\frac{i{\Slash{D}}_s}{2m_b}h_v.
\end{eqnarray}

We now compute all the time-ordered products including insertions 
of sub-leading Lagrangians that follow from the generic expressions 
(\ref{eq:combos}) and the explicit form of the Lagrangian 
(\ref{eq:lagrangians}--\ref{lhqet}) and the currents 
(\ref{eq:currents}). The diagrammatic representation of this
computation is shown in Figure~\ref{fig:tree}.\footnote{The last diagram 
in Figure~\ref{fig:tree} 
is not reproduced by a correlation function of effective
currents in SCET but requires an additional local term that we have 
not discussed. Since this term has no discontinuity it does not
contribute to the hadronic tensor and decay distributions.}
The final result is rather compact, but 
the intermediate expressions among which many cancellations take place 
are lengthy. To display this result, we introduce a short-hand 
notation 
\begin{equation}
J_a^\dagger \,J_b \,i{\cal L}_c\equiv 
i\int d^4 x\,e^{i px}\,T\left\{J_a^\dagger(x) J_b(0) \,i\int d^4 z 
\,{\cal L}_c(z)\right\}
\end{equation}
and similarly for the other terms. We also define the integral
operators ($x_+ = (n_+ x)/2$, $p_\perp=0$, $n_+ p>0$)
\begin{eqnarray}
I_2^* \,f(x_+) &\equiv& -\int d^4 x\,e^{i p x}\int\frac{d^4 k}{(2\pi)^4} 
\,e^{-i k x}\,\frac{n_+ k}{k^2+i\epsilon}\,f(x_+) 
\nonumber\\
&=& -\int d x_+ e^{i n_- p x_+}\int\frac{d n_- k}{2\pi}\,
e^{-i n_- k x_+}\,\frac{1}{n_- k+i\epsilon}\,f(x_+)
\nonumber\\
&=& i\int^{\infty}_0 d x_+\,e^{i n_- p x_+}\,f(x_+),
\\
I_3^* \,f(x_+,z_+) &\equiv&  -\int d^4 xd^4 z\,e^{i p x}
\int\frac{d^4 k}{(2\pi)^4}\frac{d^4 k^\prime}{(2\pi)^4} 
\,e^{-i k (x-z)}\,e^{-i k^\prime z}\,
\frac{n_+ k}{(k^2+i\epsilon)(k^{\prime\,2}+i\epsilon)}\,f(x_+,z_+) 
\nonumber\\
&=& I_2^* \,\frac{-i}{n_+ p}\int^{x_+}_0 d z_+\,f(x_+,z_+), 
\end{eqnarray}
and present the intermediate 
result in light-cone gauge $n_- A_s=0$ ($Y_s=1$). With this 
notation, the leading-power result for $T^{\mu\nu}$ reads in 
the tree approximation
\begin{equation}
J^{(0)\dagger}\,J^{(0)} = I_2^*\, 
\bar h_v(x_-)\Gamma_i^\dagger \frac{\slash{n}_-}{2}\Gamma_j h_v(0).
\end{equation}

Following the classification of time-ordered products 
in (\ref{eq:combos}), we calculate the individual contributions and 
obtain 
\begin{eqnarray}
&& \mbox{(I)} = J_1^{(2)\dagger} J^{(0)} = 0,
\nonumber\\
%%%
&& \mbox{(II)} = J_2^{(2)\dagger} J^{(0)} = -I_2^*\, \frac{1}{k^2}\,
(\bar h_v (-i\overleftarrow{D}_{s\perp})^2)(x_-)
\Gamma_i^\dagger \frac{\slash{n}_-}{2}\Gamma_j h_v(0),
\nonumber\\
%%%
&& \mbox{(III)} = J_3^{(2)\dagger} J^{(0)} = -I_2^*\, \frac{1}{n_+ p}\,
(\bar h_v (-i\overleftarrow{D}_s^{\mu_\perp}))(x_-)
\Gamma_i^\dagger  \frac{\slash{n}_+}{2}\gamma_{\mu_\perp}
\!\frac{\slash{n}_-}{2}\Gamma_j h_v(0),
\nonumber\\
%%%
&& \mbox{(IV)} = J_4^{(2)\dagger} J^{(0)}+J^{(0)\dagger} J_4^{(2)} = 
 \frac{1}{2 m_b} \,I_2^*\,\bigg(
(\bar h_v (-i\overleftarrow{\Slash{D}}_s))(x_-)\Gamma_i^\dagger 
\frac{\slash{n}_-}{2}\Gamma_j h_v(0) \nonumber\\
&&\hspace*{1cm} +\, 
\bar h_v(x_-)\Gamma_i^\dagger \frac{\slash{n}_-}{2}\Gamma_j 
(i\Slash{D}_s h_v)(0)\bigg),
\nonumber\\
%%%
&& \mbox{(V)} = J_1^{(1)\dagger} J_2^{(1)} = -I_2^*\, \frac{1}{n_+ p}\,
(\bar h_v (-i\overleftarrow{D}_s^{\mu_\perp}))(x_-)
\Gamma_i^\dagger  \frac{\slash{n}_-}{2}\gamma_{\mu_\perp}\!
\frac{\slash{n}_+}{2}\Gamma_j h_v(0),
\nonumber\\
%%%
&& \mbox{(VI)} = J_2^{(1)\dagger} J_2^{(1)} = 0,
\nonumber\\
%%%
&& \mbox{(VII)} = J_1^{(1)\dagger} J^{(0)}\,i{\cal L}_\xi^{(1)} = 
-I_3^*\,2 \,(\bar h_v (-i\overleftarrow{D}_s^{\mu_\perp}))(x_-)
\Gamma_i^\dagger \frac{\slash{n}_-}{2}\Gamma_j
gA_{\mu_\perp}^s(z_-)h_v(0) 
\nonumber\\
&&\hspace*{1cm} + \,I_2^*\,\frac{2}{k^2} \,(\bar h_v
(-i\overleftarrow{D}_s^{\mu_\perp}) gA_{\mu_\perp}^s)(x_-)
\Gamma_i^\dagger \frac{\slash{n}_-}{2}\Gamma_j h_v(0), 
\nonumber\\
%%%
&& \mbox{(VIII)} = (J_2^{(1)\dagger} J^{(0)}+J^{(0)\dagger} J_2^{(1)}) 
\,i{\cal L}_\xi^{(1)}= I_2^*\,\frac{1}{n_+ p}
\bigg((\bar h_v A_s^{\mu_\perp})(x_-)\Gamma_i^\dagger 
\frac{\slash{n}_-}{2}\gamma_{\mu_\perp}\!
\frac{\slash{n}_+}{2}\Gamma_j h_v(0) \nonumber\\
&&\hspace*{1cm} -\, 
\bar h_v(x_-)\Gamma_i^\dagger
\frac{\slash{n}_-}{2} \gamma_{\mu_\perp}\!\frac{\slash{n}_+}{2}\Gamma_j 
(A_s^{\mu_\perp} \!h_v)(0)\bigg),
\nonumber\\
%%%
&& \mbox{(IX)} = J^{(0)\dagger} J^{(0)}\,i{\cal L}_{\rm HQET}^{(2)} = 
I_2^*\, T\Big\{
\bar h_v(x_-)\Gamma_i^\dagger \frac{\slash{n}_-}{2}\Gamma_j h_v(0)
\,i\int d^4 z\,{\cal L}_{\rm HQET}^{(2)}(z)\Big\},
\nonumber\\
%%%
&& \mbox{(X)} = J^{(0)\dagger} J^{(0)}\,i{\cal L}_{1\xi}^{(2)} = 0,
\nonumber\\
%%%
&& \mbox{(XI)} = J^{(0)\dagger} J^{(0)}\,i{\cal L}_{2\xi}^{(2)} = 
- I_3^*\, \frac{n_+ p}{2}\,z^{\mu_\perp} z^{\lambda_\perp}\,
\bar h_v(x_-) \Gamma_i^\dagger \frac{\slash{n}_-}{2}\Gamma_j 
[D^s_{\lambda_\perp}, n_-^\nu g F^s_{\mu_\perp\nu}](z_-) 
h_v(0),
\nonumber\\
%%%
&& \mbox{(XII)} = J^{(0)\dagger} J^{(0)}\,i{\cal L}_{3\xi}^{(2)} 
= I_3^* \, \frac{1}{2}\,\bar h_v(x_-) \Gamma_i^\dagger
\frac{\slash{n}_-}{2}\gamma_{\nu_\perp} \gamma_{\mu_\perp}
i g F^s_{\mu_\perp\nu_\perp}(z_-)  \Gamma_j h_v(0),
\nonumber\\
%%%
&& \mbox{(XIII)} = \frac{1}{2} \,
J^{(0)\dagger} J^{(0)}\,i{\cal L}_{\xi}^{(1)}\,i{\cal
  L}_{\xi}^{(1)} =  - J^{(0)\dagger} J^{(0)}\,i{\cal L}_{2\xi}^{(2)}
\nonumber\\
&&\hspace*{1cm}+ \,I_3^*\,2 \,(\bar h_v g A^{\mu_\perp}_s)(x_-)
\Gamma_i^\dagger \frac{\slash{n}_-}{2}\Gamma_j
gA_{\mu_\perp}^s(z_-)h_v(0)   
\nonumber\\
&&\hspace*{1cm} - \,I_2^*\,\frac{2}{k^2} \,(\bar h_v
g A^{\mu_\perp}_s gA_{\mu_\perp}^s)(x_-)
\Gamma_i^\dagger \frac{\slash{n}_-}{2}\Gamma_j h_v(0)
\nonumber\\
&&\hspace*{1cm} - \,I_3^* \,\bar h_v(x_-)
\Gamma_i^\dagger \frac{\slash{n}_-}{2}\Gamma_j 
(gA^{\mu_\perp}_s gA_{\mu_\perp}^s - [i\partial^{\mu_\perp}
gA_{\mu_\perp}^s])(z_-)h_v(0) 
\nonumber\\  
&&\hspace*{1cm} + \,I_2^* \,\frac{1}{k^2}\,(\bar h_v 
(gA^{\mu_\perp}_s gA_{\mu_\perp}^s - [i\partial^{\mu_\perp}\!
gA_{\mu_\perp}^s]))(x_-)
\Gamma_i^\dagger \frac{\slash{n}_-}{2}\Gamma_j 
h_v(0), 
\nonumber\\  
&& \mbox{(XIV)} = J^{(0)\dagger} J^{(0)}\,i{\cal L}_{\xi q}^{(1)}\,i{{\cal
  L}_{\xi q}^{(1)}}
\nonumber\\
&&\hspace*{1cm} = \, \int d^4 x d^4 z_1 d^4 z_2 \,e^{i p x} 
\int \frac{d^4k}{(2\pi)^4}  \frac{d^4k_1}{(2\pi)^4}  \frac{d^4k_2}{(2\pi)^4}
\,e^{-i k_1(x-z_2)}\,e^{-i k(z_2-z_1)}\,e^{-i k_2 z_1}
\nonumber\\
&&\hspace*{2cm} \frac{n_+ k_1\,n_+ k_2}{k^2 k_1^2 k_2^2} \,
g^2\,\bar h_v(x_-) \Gamma_i^\dagger \frac{\slash{n}_-}{2} \gamma^{\mu_\perp}\!
T^A q_s(z_{2-})\,\bar q_s(z_{1-}) T^A
\gamma_{\mu_\perp}\!\frac{\slash{n}_-}{2}\Gamma_j h_v(0)
\\
&&\hspace*{1cm} 
=\,I_2^* \,\frac{1}{n_+ p} \,
\int^{x_+}_0 \!\!dz_{1+} \!\int^{x_+}_{z_{1+}} \!\!d z_{2+} \,
g^2\,\bar h_v(x_-) \Gamma_i^\dagger \frac{\slash{n}_-}{2} 
\gamma^{\mu_\perp}\!
T^A q_s(z_{2-})\,\bar q_s(z_{1-}) T^A
\gamma_{\mu_\perp}\!\frac{\slash{n}_-}{2}\Gamma_j h_v(0). 
\nonumber
\end{eqnarray}
In order to arrive at this result we have made repeated use of the 
equation of motion (\ref{freeeom}) for the collinear propagator. The 
expressions simplify considerably after adding (III)+(V)+(VIII) and 
(II)+(VII)+(XI)+(XII)+(XIII). Together with the other non-vanishing 
terms (IV), (IX) and (XIV) we obtain for the current correlation
function at order $1/m_b$,
\begin{eqnarray}
&& T_{\rm tree}^{1/m_b} = -\int\frac{d x_+ d\omega}{2\pi} \,
e^{i (n_- p-\omega) x_+}\,\frac{1}{\omega+i\epsilon}\,\times
\bigg\{
\nonumber\\
&& \hspace*{1cm}
T\Big\{(\bar h_v Y)(x_-)\Gamma_i^\dagger \frac{\slash{n}_-}{2}\Gamma_j 
(Y^\dagger h_v)(0)\,
\,i\int d^4 z\,{\cal L}_{\rm HQET}^{(2)}(z)\Big\}
\nonumber\\
&&  \hspace*{1cm}
+\,\frac{1}{2 m_b} \Big[
(\bar h_v (-i\overleftarrow{\Slash{D}}_s)Y)(x_-)\Gamma_i^\dagger 
\frac{\slash{n}_-}{2}\Gamma_j (Y^\dagger h_v)(0) + 
(\bar h_v Y)(x_-)\Gamma_i^\dagger \frac{\slash{n}_-}{2}\Gamma_j 
(Y^\dagger i\Slash{D}_s h_v)(0)\Big]
\nonumber\\
&&  \hspace*{1cm} -\, \frac{1}{n_+ p}\,
\Big[(\bar h_v  (-i\overleftarrow{D}_s^{\mu_\perp})Y)(x_-)\Gamma_i^\dagger 
\frac{\slash{n}_+}{2}\gamma_{\mu_\perp}\!
\frac{\slash{n}_-}{2}\Gamma_j (Y^\dagger h_v)(0) \nonumber\\[-0.2cm]
&&\hspace*{4cm} +\, 
(\bar h_v Y)(x_-)\Gamma_i^\dagger
\frac{\slash{n}_-}{2} \gamma_{\mu_\perp}\!\frac{\slash{n}_+}{2}\Gamma_j 
(Y^\dagger (-i \overleftarrow{D}_s^{\mu_\perp})h_v)(0)\Big]
\nonumber\\
&&  \hspace*{1cm} +\, \frac{i}{n_+ p}\,
\int^{x_+}_0 \!d z_+(\bar h_v Y)(x_-)\Gamma_i^\dagger \frac{\slash{n}_-}{2}
(Y^\dagger (-i \overleftarrow{\Slash{D}}_{s\perp}) 
(-i \overleftarrow{\Slash{D}}_{s\perp})  Y)(z_-)\Gamma_j (Y^\dagger h_v)(0)
\nonumber\\
&&  \hspace*{1cm} +\, \frac{1}{n_+ p}\,
\int^{x_+}_0 \!dz_{1+} \int^{x_+}_{z_{1+}} d z_{2+} \, g^2\,
(\bar h_v Y)(x_-)\Gamma_i^\dagger \frac{\slash{n}_-}{2} 
\gamma^{\mu_\perp}\! T^A (Y^\dagger q)(z_{2-})\,
 \nonumber\\[-0.2cm]
&&\hspace*{4cm} \times \,(\bar q Y)(z_{1-}) T^A
\gamma_{\mu_\perp}\!\frac{\slash{n}_-}{2}\Gamma_j (Y^\dagger h_v)(0) 
\bigg\}.
\label{final1}
\end{eqnarray}
We have reinserted the soft Wilson lines $Y$, which makes this expression 
gauge invariant. The derivatives are understood to
act on everything to their right (or to their left when indicated by
the arrow) independent of the position argument of the field.
It is worth noting that the double insertion of ${\cal L}_\xi^{(1)}$ 
(XIII) was expected to be tetra-local. However, only the 
delta-function terms survive in the application of the 
equation of motion identity and the complexity of this term is reduced
to a tri-local term of the form $\bar h_v(x_-) [iD_\perp iD_\perp](z_-)
h_v(0)$. On the other hand, the four-quark contribution from the 
double insertion of  ${\cal L}_{\xi q}^{(1)}$ {\em is} tetra-local. 

We have also performed the calculation in a general frame where
$p_\perp\not =0$, in which the result is a significantly more  
complicated expression. In particular, the double insertion 
of ${\cal L}_\xi^{(1)}$ gives a tetra-local term proportional to 
$p_\perp^2$. Superficially, this seems to require a much larger set of
shape-functions, including a different degree of locality. However, 
the hadronic tensor depends only on two kinematic invariants $vp$ and 
$p^2$, so two out of the three variables $n_+ p$, $n_- p$ and 
$p_\perp^2$ must be sufficient to reconstruct the complete
information. We therefore conclude that the specific convolutions 
of jet-functions and tetra-local shape-functions from the double
insertion of ${\cal L}_\xi^{(1)}$ (and other similar terms not present
for $p_\perp=0$) in the general frame cannot
contain independent non-perturbative information despite their
appearance. We can see this technically by noting that the transverse
momentum is defined with respect to a choice of vectors $n_-$, $v$. 
A transverse Lorentz boost from a frame with $p_\perp=0$ to a frame 
with $p_\perp\sim \lambda$ can be effected by a reparameterization 
of $n_-\to n_- +2\epsilon_\perp-\epsilon_\perp^2 n_+$ with 
$\epsilon_\perp \sim -p_\perp/n_+ p$ 
and $v$ fixed. The complete SCET expansion is invariant, because it 
reproduces a Lorentz-invariant theory, but the transformation
reshuffles terms in the $\lambda$ expansion. In particular, the 
leading-order Lagrangian changes ${\cal L}^{(0)} \to {\cal L}^{(0)} 
+\epsilon_\perp  \delta{\cal L}^{(0)}+\ldots$. Accordingly for matrix
elements 
\begin{equation}
\langle O\rangle \to 
 \langle O\rangle + \epsilon_\perp  \int d^4z \,\langle T\big\{O \,i  
\delta{\cal L}^{(0)}(z)\big\} \rangle + \ldots,
\end{equation}
from which it is clear that the leading-power 
bi-local term in the frame $p_\perp=0$ gives rise to tetra-local terms
proportional to $p_\perp^2$ at order $\lambda^2$ in the general
frame. While this explains the structure of terms in the general frame,
we did not verify explicitly the equivalence of the expressions, which
seems to be a technical but unilluminating task. 

We now proceed to the evaluation of the hadronic tensor 
(\ref{hadronic_tensor}). Starting from (\ref{final1}) this requires 
that we i) take the imaginary part, ii) take 
the  $\bar B$ meson matrix element and iii) insert the 
decomposition of the soft matrix elements into scalar shape-functions 
as given in Section~\ref{sec:basis}.  The imaginary 
part of  (\ref{final1}) is obtained by the replacement 
\begin{equation}
\frac{1}{\omega+i\epsilon}\to -\pi \delta(\omega).
\end{equation}
The $\bar B$ meson matrix elements can all be expressed 
in terms of the previously defined shape-functions, if the 
relation 
\begin{eqnarray}
&&\langle \bar B|(\bar h_v Y)(x_-)_{a\alpha} 
 (Y^\dagger i D_s^{\mu} Y)(z_-)_{cd} 
 (Y^\dagger h_v)(0)_{b\beta}|\bar B\rangle 
\nonumber\\
&&\hspace*{2cm}= \,
\langle \bar B|(\bar h_v Y)(-x_-)_{b\beta} 
 (Y^\dagger (-i) \overleftarrow{D}_s^{\mu} Y)(z_--x_-)_{dc} 
 (Y^\dagger h_v)(0)_{a\alpha}|\bar B\rangle^*  
\end{eqnarray}
and $\partial_{\mu_\perp}\langle \bar B|\ldots |\bar B\rangle=0$ 
is used. 
The scalar decomposition contains odd terms proportional to
$\epsilon^\perp_{\mu\nu}$, which can be eliminated using 
$\epsilon^\perp_{\mu\nu} \gamma_\perp^\nu\gamma_5 = 
\gamma_{\mu_\perp} (\slash{v}\slash{n}_--\slash{n}_-\slash{v})/2$. 
We then express the position space shape-functions in terms of 
their Fourier transforms defined as in (\ref{defshape}) and 
perform the integrations over the positions left in (\ref{final1}). 
The result for the hadronic tensor including now the leading-power
contribution (in the tree approximation) becomes 
\bea
&& W^{\mu\nu}=\int d\omega \,\delta(n_- p - \omega)\,\Bigg\{
\nonumber\\ 
&& \hspace*{0.5cm}
\frac{1}{2}{\rm tr}\Big[P_+\Gamma_i^\dagger\frac{\slash{n}_-}{2}\Gamma_j\Big]
\left(S(\omega)+\frac{1}{2 m_b} 
\left[s_{\rm kin}(\omega)+C_{\rm mag}(m_b/\mu) s_{\rm mag}(\omega)\right]
-\frac{u_s(\omega)+2 v_s(\omega)}{\np p}\right)\nl
&& \hspace*{0.5cm}+\,
\frac{1}{4 m_b} 
 {\rm tr}\Big[(\slash{n}_- -\slash{v})\Gamma_i^\dagger
\frac{\slash{n}_-}{2}\Gamma_j\Big]
\left(\omega S(\omega)-t(\omega)\right)
 +
\frac{1}{4\np p}{\rm tr} \Big[P_+\gamma_\perp^\alpha\gamma_5 
\Gamma_i^\dagger\gamma_{\alpha\perp}\gamma_5 
\Gamma_j\Big] \,t(\omega)
\nl
&& \hspace*{0.5cm}+\,
\frac{1}{2\np p}
{\rm tr}\Big[P_+ \slash{n}_-\gamma_5P_+
\Gamma_i^\dagger\frac{\slash{n}_-}{2}\gamma_5 \Gamma_j\Big] 
\,(u_a(\omega)-2 v_a(\omega))\Bigg\},
\label{final2}
\eea  
where $P_+=(1+\slash{v})/2$, and we introduced the definitions 
\begin{eqnarray}
u_s(\omega)&=& \int d\omega_1 d\omega_2 \,J_2(\omega;\omega_1,\omega_2) 
\,(u_1(\omega_1)\delta(\omega_2)+U_1(\omega_1,\omega_2)),
\nonumber\\
u_a(\omega)&=& \int d\omega_1 d\omega_2 \,J_2(\omega;\omega_1,\omega_2) 
\,U_3(\omega_1,\omega_2),
\nonumber\\
v_s(\omega)&=& \int d\omega_1 d\omega_2 d\omega_3 \,
J_3(\omega;\omega_1,\omega_3,\omega_2) 
\,g^2\,{\cal V}_{10}(\omega_1,\omega_2,\omega_3),
\nonumber\\
v_a(\omega)&=& \int d\omega_1 d\omega_2 d\omega_3 \,
J_3(\omega;\omega_1,\omega_3,\omega_2) 
\,g^2\,{\cal V}_{13}(\omega_1,\omega_2,\omega_3),
\label{effshape}
\end{eqnarray}
\begin{eqnarray}
J_2(n_- p;\omega_1,\omega_2)&=&
-\frac{1}{\pi}\,{\rm Im}\frac{(\np p)^2}{p_{\omega_1}^2 p_{\omega_{12}}^2}
=\frac{1}{\omega_2}\big(
\delta(\nm p -\omega_1-\omega_2)-\delta(\nm p-\omega_1)\big),
\nonumber\\
J_3(n_- p;\omega_1,\omega_2,\omega_3)&=&
-\frac{1}{\pi}\,
{\rm Im}\frac{(\np p)^3}{p_{\omega_1}^2 p_{\omega_{12}}^2 
p_{\omega_{123}}^2}
\nonumber\\
&& \hspace*{-2cm}
=\,\frac{\delta(n_- p-\omega_1)}{\omega_2 (\omega_2+\omega_3)} - 
 \frac{\delta(n_- p-\omega_1-\omega_2)}{\omega_2 \omega_3} + 
 \frac{\delta(n_- p-\omega_1-\omega_2-\omega_3)}
 {\omega_3 (\omega_2+\omega_3)} 
\nonumber\\[0.2cm]
&& \hspace*{-1.4cm}
-\pi^2\delta(n_- p-\omega_1)\delta(\omega_2)\delta(\omega_3),
\end{eqnarray}
and used that $t(\omega)$ is real \cite{BLM}. The denominators in the 
definition of $J_2$ and $J_3$ are understood to be supplied with a
principal value definition. Eq.~(\ref{final2}) is our final result, 
valid for arbitrary Dirac structures $\Gamma_i$ and $\Gamma_j$ of 
the weak currents. Despite the fact that the hadronic matrix elements 
are tri- and tetra-local the final result can be written as
single convolutions of integrated shape-functions. Hence, from a 
phenomenological point of view, the $1/m_b$ corrections can be
parameterized by a set of functions depending only on a single
variable, just as at leading power. We discuss in 
Section~\ref{sec:beyondtree} whether this holds when loop corrections
are included. The factor $g^2$ in the definition of the integrated 
four-quark shape-functions $v_{s,a}(\omega)$ should not lead to the 
conclusion that these contributions are suppressed. In fact, the same 
factor of $g^2$ is also present in the tetra-local piece of the 
two-derivative matrix element (\ref{twoDshape}), which is 
$g^2\bar h_v A_\perp A_\perp h_v$ in light-cone gauge, and here the 
$g^2$ is conventionally assumed to be normalized at the non-perturbative 
strong interaction scale. 

The $\bar B$ decay distributions are most conveniently expressed in terms 
of the scalar components of the hadronic tensor. In the following we
give the result for the currents relevant to semi-leptonic 
$\bar B\to X_u \ell\bar \nu$ decay and the radiative 
$\bar B\to X_s \gamma$ decay in the convention specified by 
(\ref{hadtensordef}). For the semi-leptonic decay
\begin{equation}
\Gamma_i^\dagger=\gamma^\mu(1-\gamma_5),\quad 
\Gamma_j=\gamma^\nu(1-\gamma_5),
\end{equation}
and we find 
\bea
W_1&=&\frac{2}{\np p} \int d \omega \,\delta(n_- p-\omega)
\bigg[\left(1+\frac{\nm p}{\np p}\right)S(\omega) + 
\frac{s_{\rm kin}(\omega)+C_{\rm mag}(m_b/\mu) s_{\rm mag}(\omega)}{2 m_b}
\nonumber\nl 
&& -\, \frac{
\omega S(\omega)-t(\omega)}{m_b}
-\frac{u_s(\omega)+u_a(\omega)}{\np p} 
-\frac{2 ({v}_s(\omega)- {v}_a(\omega))}{\np p}\bigg],
\nonumber\\
W_2 &=&\frac{1}{2} W_3 = 
-\frac{2 \nm p}{\np p}\int d\omega \,\delta(n_- p-\omega)\,S(\omega),
\nonumber\\
W_4&=&-\frac{4}{(\np p)^2}\int d\omega  \,\delta(n_- p-\omega)
\,t(\omega),
\nonumber\\
W_5&=& \frac{8}{(\np p)^2}\int d\omega\,\delta(n_- p-\omega)
\bigg[
\frac{\omega S(\omega)-t(\omega)}{m_b}+\frac{
t(\omega)+u_a(\omega)-2 v_a(\omega)}{\np p}\bigg].
\eea
The result is given in the frame $p_\perp=0$, where $vp=(n_+ p+n_- p)/2$ and 
$p^2=n_+p \,n_- p$. This can be used to convert the expressions to hadronic 
variables, see (\ref{hadvar}). 
We recall that the expressions are valid in the kinematic region 
where $n_+ p\sim m_b$ and $n_- p\sim \Lambda_{\rm QCD}$. Note that only $W_1$ 
has a leading-power term. 

Turning to the radiative decay $\bar B\to X_s\gamma$, we note that 
the photon momentum is $q=E_\gamma n_+$ in the frame 
$p_\perp=-q_\perp=0$. Neglecting terms proportional to $q^\mu$ which 
vanish when the hadronic tensor is contracted with the photon 
polarization vector, we can take the current to be 
\begin{equation}
\Gamma_i^\dagger=\frac{1}{4}\,[\slash{n}_+,\gamma_\perp^\mu]
(1-\gamma_5),\quad 
\Gamma_j=\frac{1}{4}\,[\gamma_\perp^\nu,\slash{n}_+](1+\gamma_5).
\end{equation}
It is simpler here not to decompose the hadronic tensor into scalar 
functions. Evaluating the traces in (\ref{final2}) we obtain 
\bea
W^{\mu\nu}&=&-(g_\perp^{\mu\nu}+\eps_\perp^{\mu\nu})
\int d\omega \,\delta(n_-p-\omega)\Bigg\{S(\omega) + 
\frac{s_{\rm kin}(\omega)+C_{\rm mag}(m_b/\mu) s_{\rm mag}(\omega)}{2 m_b}
\nonumber\\
&&+\,\frac{\omega S(\omega)-t(\omega)}{m_b}
-\frac{u_s(\omega)+u_a(\omega)}{\np p} 
-\frac{2 ({v}_s(\omega)- {v}_a(\omega))}{\np p}\Bigg\}.
\eea
When the polarization of the photon is not observed, the 
$\eps_\perp^{\mu\nu}$ term does not contribute and the photon 
energy spectrum reads
\bea 
\frac{1}{2\Gamma}\,\frac{d\Gamma}{d E_\gamma} &=& 
\left(1-\frac{2 n_- p}{m_b}\right) \,S(n_- p) + 
\frac{s_{\rm kin}(n_- p)+C_{\rm mag}(m_b/\mu) s_{\rm mag}(n_- p)}{2 m_b}
\nonumber\\
&&-\,\frac{1}{m_b} \Big(t(n_- p)+u_s(n_- p)+u_a(n_- p) 
+ 2 ({v}_s(n_- p)- {v}_a(n_- p)\Big)
\label{photonspec}
\eeq
with $n_- p=m_b-2 E_\gamma$. This result is valid at tree level, 
and in the approximation where the four-quark operators in the 
weak effective Hamiltonian are neglected.\footnote{There is no tree level 
contribution from four-quark operators at leading power. At order 
$1/m_b$ a non-zero contribution arises from soft gluons attached to 
the charm quark loop. The degree of non-locality of these terms 
depends on whether an expansion in $1/m_c$ is performed.}  

The hadronic tensor involves a power correction  
$(s_{\rm kin}(\omega)+C_{\rm mag}(m_b/\mu) s_{\rm mag}(\omega))/(2 m_b)$ 
from the insertion of the $1/m_b$ corrections to the HQET 
Lagrangian, ${\cal L}_{\rm HQET}^{(2)}$, because it is conventional to
evaluate the soft matrix element with the leading-power 
HQET Lagrangian. This is advantageous in applications of HQET,
where use is made of the heavy-quark spin-flavour symmetries. In the
present case it is more convenient to not treat the power corrections
to the HQET Lagrangian in the interaction picture. Then the
above-mentioned term should be omitted, but the matrix elements 
$\langle \bar B|\ldots |\bar B\rangle$ are evaluated with 
the exact HQET Lagrangian including ${\cal L}_{\rm HQET}^{(2)}$. 
In this picture the HQET matrix elements have a (small)
$m_b$ dependence, but this is not an issue as long as all corrections
up to a required order are included. It is
further useful to regard the shape-functions as functions of the {\em
  hadronic} variable $n_- P = n_- p +M_B-m_b$ ($P$ is the total
momentum of the hadronic final state), i.e.~for any function
$f(\omega)$ above we define $f(\omega)=\hat f(\omega+M_B-m_b)$, 
such that a physical spectrum such as (\ref{photonspec}) is expressed
in terms of $\hat f(n_- P)$. Note that $M_B-m_b$ is 
{\em not} the HQET parameter $\bar\Lambda$, but the difference 
between the physical meson mass and the heavy quark mass. 
The distinction is relevant at the level of power corrections to
the spectrum. The advantage of taking the matrix elements with respect
to the exact HQET Lagrangian is that the support of the functions 
$\hat f$ is then from 0 to $\infty$.\footnote{The upper limit is
  infinity as a consequence of factorization, which removes the
  physical upper limit and replaces it by a cut-off. However, dimensional 
  regularization does not provide a dimensionful cut-off.}

Comparing our result with previous work 
\cite{BLM,Bauer:2002yu,Burrell:2003cf}, we find agreement 
for the photon energy spectrum in the radiative decay \cite{Bauer:2002yu} 
and the hadronic invariant mass spectrum in the semi-leptonic 
decay \cite{Burrell:2003cf}, provided we neglect the 
effect from the four-quark shape-functions $v_{s,a}(\omega)$. 
However, our general result (\ref{final1}) and the lepton energy 
spectrum in the semi-leptonic decay  following from this result differ 
from \cite{Bauer:2002yu} even when the four-quark shape-functions are 
neglected. The short-distance expansion of the 
hadronic tensor was obtained in previous work by direct matching 
of QCD to heavy quark effective theory without the intermediate 
use of soft-collinear effective theory. In the tree approximation 
the two approaches should give the same result. However, some of the results 
in \cite{BLM,Bauer:2002yu} are obtained by an expansion in transverse 
momentum of a hadronic tensor that effectively includes the 
integration over neutrino momentum. This can lead to incorrect results, 
because the transverse momentum relevant to the expansion is the 
transverse momentum of partons relative to the jet, not the jet and 
the neutrino. The calculations of 
the photon energy spectrum in the radiative decay 
and of the hadronic invariant mass spectrum in the semi-leptonic 
decay are not affected by this problem. 

\section{Remarks on factorization beyond tree level}
\label{sec:beyondtree}

We have just seen that at tree level many simplifications take place
that reduce the degree of non-locality of shape-functions appearing 
in the $1/m_b$ corrections. In this section we investigate whether 
this simplification persists beyond tree level. Since the conclusion 
will be negative, it suffices to illustrate this point for the case of
abelian gauge fields. 

The manipulations below rely on the analogue of the QED Ward identity 
for the leading power (abelian) SCET Lagrangian after the collinear field
redefinition that decouples soft and collinear fields. Defining 
\begin{eqnarray}
J_+ &=& \bar\xi \,\frac{\slash{n}_+}{2}\xi,
\nonumber\\
J_\perp^\mu &=& \bar\xi \left(i\overleftarrow{\Slash{D}}_{\perp c}
\frac{1}{i n_+ \overleftarrow{D}_c}
\,\gamma^{\mu_\perp}+\gamma^{\mu_\perp}\frac{1}{i n_+ D_c}
i\Slash{D}_{\perp c}
\right)\frac{\slash{n}_+}{2}\xi, 
\nonumber \\ 
J_- &=& \bar\xi i\overleftarrow{\Slash{D}}_{\perp c}\frac{1}{i n_+
  \overleftarrow{D}_c}\frac{\slash{n}_+}{2}
\frac{1}{i n_+ D_c}i\Slash{D}_{\perp c}\xi,
\end{eqnarray}
the Ward identity for jet-functions reads
\begin{eqnarray}
&& n_-\partial_{(z)}
\langle 0|T\Big\{J_+(z) \chi(x_1) 
\ldots \chi(x_m)\Big\}|0\rangle + 
\partial_{(z)}^{\mu_\perp}
\langle 0|T\Big\{J_{\perp\mu}(z) \chi(x_1) 
\ldots \chi(x_m)\Big\}|0\rangle 
\nonumber\\
&& \hspace*{1cm} 
+ n_+\partial_{(z)}
\langle 0|T\Big\{J_-(z) \chi(x_1) 
\ldots \chi(x_m)\Big\}|0\rangle 
\nonumber\\
&& =  
- \sum_{k=1}^m (-1)^P \,\delta^{(4)}(z-x_k) 
\,\langle 0|T\Big\{\chi(x_1) 
\ldots \chi(x_m)\Big\}|0\rangle,
\end{eqnarray}
where $P=0$ if $\chi=\xi$ and $P=1$ if $\chi=\bar\xi$. 
This generalizes to insertions of composite operators by taking the 
local limit of a product of fundamental fields. (For derivative
operators some of the delta-functions terms will then be derivatives
of delta-functions.) Our main application of the Ward identity will be
in the form
\begin{eqnarray}
&& \int d^4 z\,f(z_+,z_\perp)\,n_-\partial_{(z)}
\langle 0|T\Big\{(\bar\xi\,\frac{\slash{n}_+}{2}\xi)(z) \chi(x_1) 
\ldots \chi(x_m)\Big\}|0\rangle 
\nonumber\\
&&\hspace*{2cm}
= \int d^4 z\,[\partial_{\mu_\perp} f(z_+,z_\perp)]\,
\langle 0|T\Big\{J_\perp^\mu(z) \chi(x_1) 
\ldots \chi(x_m)\Big\}|0\rangle 
\nonumber\\
&&\hspace*{2cm}
- \sum_{k=1}^m (-1)^P \,f(x_{k+},x_{k\perp}) 
\,\langle 0|T\Big\{\chi(x_1) 
\ldots \chi(x_m)\Big\}|0\rangle.
\end{eqnarray}

\paragraph{\it 1. Degree of multi-locality.} 
The highest degree of non-locality in the product of soft fields 
at order $1/m_b$ comes from the double insertions of 
${\cal L}^{(1)}_\xi$ and ${\cal L}^{(1)}_{\xi q}$. The double
insertion of the mixed collinear-soft quark Lagrangian
${\cal L}^{(1)}_{\xi q}$ is tetra-local even at tree level. On the
other hand 
\begin{equation}
\langle \bar B | 
T\Big\{(\bar h_v\Gamma_i^\dagger W_c^\dagger\xi)(x) 
(\bar\xi W_c\Gamma_j h_v)(0) 
\,\frac{1}{2} \,i\!\int d^4 z_1 \,{\cal L}_\xi^{(1)}(z_1) 
\, i\!\int d^4 z_2 \,{\cal L}_\xi^{(1)}(z_2)\Big\}|\bar B\rangle,
\label{tetra}
\end{equation}
while superficially tetra-local, actually reduces to a tri-local 
term at tree level. Does this hold to all orders?

In the abelian theory the collinear Wilson lines in 
${\cal L}^{(1)}_\xi$ drop out. We use (\ref{fieldtocov}) and 
integrate by parts to write (after the collinear-soft decoupling field
redefinition)
\begin{equation}
\label{nmterm}
i\!\int d^4 z \,{\cal L}_\xi^{(1)}(z) = 
\int d^4 z \,z^{\mu_\perp} \,(Y^\dagger [i D^s_{\mu_\perp}Y])(z_-) \,
i n_-\partial_{(z)} (\bar\xi\,\frac{\slash{n}_+}{2}\xi)(z).
\end{equation}
Applying the Ward identity twice to (\ref{tetra}), we can 
cast this contribution into the form 
\begin{eqnarray}
&& \!-\frac{1}{2} \int d^4 z_1\int d^4 z_2\,
\langle \bar B | (\bar h_v Y\Gamma_i^\dagger)(x_-)_\alpha 
(Y^\dagger [i D^s_{\mu_\perp}Y])(z_{1-})
(Y^\dagger [i D^s_{\nu_\perp}Y])(z_{2-})
(\Gamma_j Y^\dagger h_v)(0)_\beta |\bar B\rangle
\nonumber\\
&&\hspace*{1cm}\,\times 
\langle 0|T\Big\{(W_c^\dagger\xi)(x)_\alpha (\bar\xi W_c)(0)_\beta\,
J_\perp^\mu(z_1)\,J_\perp^\nu(z_2)
\Big\}|0\rangle + \mbox{$\delta$-function terms},
\end{eqnarray}
where ``$\delta$-function terms'' refers to contributions from 
the Ward identity which are only bi- or tri-local. In this form 
it is evident that the tetra-local term vanishes at tree level,  
because the jet-function with two insertions of the transverse current 
integrated over the transverse positions vanishes in this
approximation. In momentum space, the derivative in the definition of 
$J_\perp^\mu$ must be proportional to the transverse component of a
collinear vector, but in the frame where $p_\perp=0$ there is no such
vector. By the same line of reasoning the jet-function cannot be
expected to be zero beyond tree level. When the insertion of the
current appears inside a collinear loop diagram, the transverse 
derivative produces factors of transverse loop momentum and an 
even number of such factors results in a $g_\perp^{\mu\nu}$ 
term after integration. We therefore conclude that beyond tree level 
tetra-local terms appear also from the double insertions of the 
sub-leading collinear Lagrangian, and the complexity of the expression
is not reduced.

\paragraph{\it 2. Transformation to the covariant derivative basis.} 
The SCET Lagrangian has many terms in which the combination 
$x_\perp^\mu n_-^\nu g F_{\mu\nu}(x_-)$ occurs. These can be 
removed by applying the identity (\ref{fieldtocov}) (see 
also (\ref{nmterm})). At tree level 
we have seen that the various terms organize themselves somewhat
miraculously into covariant derivatives $(Y^ \dagger iD_s^\mu Y)(z)$ 
despite the fact that this object represents two separate 
entities (see the two terms on the right hand side of 
(\ref{covD})). For this reason we have chosen the covariant derivative
basis for the soft matrix elements in Section~\ref{sec:basis}.

Using the Ward identity it is straightforward to show that 
at order $\lambda$, the two terms 
$ J_1^{(1)\dagger} J^{(0)}$ and 
$J^{(0)\dagger} J^{(0)}\,i{\cal L}_\xi^{(1)}$ can be combined 
into an expression containing a soft matrix element with a single 
covariant derivative (as in (\ref{oneDshape})). At order $\lambda^2$,  
we find (using the short-hand notation of Section~\ref{sec:tree}) 
\begin{eqnarray}
&&\mbox{(II)} +  \mbox{(VII)} +  \mbox{(XI)} +  \mbox{(XII)} 
\nonumber\\
&&\hspace*{1cm} =\,
J_2^{(2)\dagger} J^{(0)} 
+ J_1^{(1)\dagger} J^{(0)}\,i{\cal L}_\xi^{(1)}
+ J^{(0)\dagger} J^{(0)}\,i{\cal L}_{2\xi}^{(2)} 
+ J^{(0)\dagger} J^{(0)}\,i{\cal L}_{\xi}^{(1)}\,i{\cal
  L}_{\xi}^{(1)}
\nonumber\\
&& \hspace*{1cm} =\,\frac{1}{4}{\rm tr}
\Big[P_+\Gamma_i^\dagger\frac{\slash{n}_-}{2}
\Gamma_j\Big]\,i\int d^4 x \,e^{i p x}\,\times\bigg\{
\nonumber\\
&& \hspace*{1.5cm} \int d^4 z_1\int d^4 z_2\,
\langle \bar B | (\bar h_v Y)(x_-)
(Y^\dagger i D^s_{\mu_\perp}Y)(z_{1-})
(Y^\dagger i D^s_{\nu_\perp}Y)(z_{2-})
(Y^\dagger h_v)(0)|\bar B\rangle
\nonumber\\
&&\hspace*{2.5cm}\,\times 
\langle 0|T\Big\{(\bar\xi W_c)(0)(W_c^\dagger\xi)(x)\,
J_\perp^\mu(z_1)\,J_\perp^\nu(z_2)
\Big\}|0\rangle
\nonumber\\
&& \hspace*{1.5cm} -\,2i \int d^4 z\,
\langle \bar B | (\bar h_v Y)(x_-)
(Y^\dagger (i D^s_{\perp})^2 \,Y)(z_{-})
(Y^\dagger h_v)(0)|\bar B\rangle
\nonumber\\
&&\hspace*{2.5cm}\,\times  
\langle 0|T\Big\{(\bar\xi W_c)(0)(W_c^\dagger\xi)(x)\,
(\bar\xi \frac{1}{i n_+D_c}
\frac{\slash{n}_+}{2}\xi)(z)
\Big\}|0\rangle+\ldots\bigg\},
\label{reorganize}
\end{eqnarray}
where the ellipses denote further terms 
involving $\partial_\perp A_\perp$. Here we recognize the 
tetra-local shape-function with two covariant derivatives 
defined in (\ref{twoDshape}). As mentioned above and seen explicitly 
in Section~\ref{sec:tree} this term is multiplied by a jet-function
that is zero at tree level in the frame $p_\perp=0$. The second 
term does not vanish at tree level and reproduces the corresponding
terms in Section~\ref{sec:tree}.
 
It therefore appears that in general the structure of terms 
that follows from the currents and Lagrangian insertions of SCET
can be systematically simplified by eliminating all the terms that
involve factors of position from the multipole expansion. 
Application of the Ward identities then organizes the terms into
covariant derivatives as displayed above. While we did not prove this
statement in general, and in particular for the non-abelian case, 
we may note that when the hybrid momentum-position space 
formulation of SCET is employed \cite{Bauer:2000yr}, we are led 
directly to an expansion in terms of covariant derivatives. 

\paragraph{\it 3. Convolutions with ``effective'' shape-functions.} 
In the tree approximation the had\-ronic tensor (\ref{final2}) 
can be written in terms of the ``effective'' shape-functions defined 
in (\ref{effshape}) that depend only on a single variable. This 
seems to be a simplification, since the information about hadronic
matrix elements is encoded in a set of single- rather than
multi-variable functions. 

This can always be done for contributions to the hadronic tensor 
from two-body currents in SCET. In these cases $n_+ p$ is the only large 
collinear momentum on which the jet-function can depend, and Lorentz 
invariance and dimensional analysis imply that the dependence 
on $n_+ p$ can be factored out except for powers of logarithms 
of $n_+ p/\mu$. The dependence on the remaining external momentum 
$n_- p$ can always be formally decoupled by introducing 
$\int d\omega \,\delta(n_- p-\omega)$ to define effective 
shape-functions that depend on only one variable $\omega$. The
so-defined effective shape-functions are in fact convolutions 
of jet-functions and soft matrix elements as exhibited in 
(\ref{effshape}) in the tree approximation. However, beyond this 
approximation there are contributions from three- and four-body
currents as given in Section~\ref{sec:currents}. The jet-functions 
are then convolutions of a number of large collinear momentum 
components of the effective vertices and the dependence on 
$n_+ p$ can no longer be factored, but arises after the convolution 
with hard coefficient functions. Even restricting attention to 
the contributions from two-body currents, the definitions
of effective shape-functions such $u_{s,a}(\omega)$ and
$v_{s,a}(\omega)$  in (\ref{effshape}) are useful only in the tree
approximation, because the weight functions (jet-functions) $J_{2,3}$ 
change at order $\alpha_s$. In any case, since the number of functions
exceeds the number of observables, one must return to their
definitions in terms of multi-local hadronic 
matrix elements in order to obtain an estimate from a particular
hadronic model. 

\section{Estimate of four-quark contributions}
\label{sec:phenom}

In this section we shall make a few remarks concerning the phenomenological 
implications of our results. We do not perform an exhaustive analysis,
rather we restrict ourselves to the new contribution from the
four-quark shape-functions, which has not been studied in  
\cite{BLM,Bauer:2002yu}. 
The contribution we are going to consider involves $\bar B$ meson matrix
elements $\langle \bar B|\bar h_v(x_-) h_v(0)\bar q(z_{1-}) q (z_{2-})|
\bar B \rangle$ of two heavy quark and two soft light quark fields, all at
different positions on the light-cone. An interesting aspect of this 
contribution is that the soft quark flavour originates from the weak decay 
vertex. For semi-leptonic $\bar B\to X_u \ell \bar \nu$ decay it has 
to be a $u$ quark. The matrix element is therefore expected to be
different for the decay of a charged $\bar B$ meson (where the flavour of
the spectator quark matches $u$) and a neutral $\bar B$ meson. As we have 
seen, this effect is of order $1/m_b$ in the decay spectra, 
while such differences in the total semi-leptonic rates between 
charged and neutral $\bar B$ mesons are at least of order $1/m_b^3$ 
\cite{Leibovich:2002ys,Voloshin:2001xi}.  

The general expression for the four-quark shape-function has been given in 
(\ref{fourquarkshape}). It involves 16 scalar functions, but 
only two of these, ${\cal V}_{10}$ and ${\cal V}_{13}$, 
appear in the hadronic tensors of the inclusive decays 
$\bar B \to X_s\gamma$ and 
$\bar B \to X_u  \ell \bar{\nu}$ at tree level. These functions are 
tetra-local and their Fourier transforms depend on three 
variables. However, the result can be written in terms of two
effective shape-functions $v_{s,a}(\omega)$, see (\ref{effshape}).

In order to obtain a quantitative estimate of these contributions 
we have to employ a model, which at present can be only quite
crude. Matrix elements of four-quark operators are often approximated 
in the ``vacuum saturation approximation,'' where 
(leaving out spin and colour indices)
\begin{equation} 
\label {Assume}
\langle \bar B | \bar{h}_v (x_-) u (z_{2-})\bar{u} (z_{1-}) h_v  (0) 
                | \bar B \rangle    
\quad \to  \quad 
  \langle \bar B | \bar{h}_v (x_-) u(z_{2-})  | 0 \rangle
              \langle 0 |  \bar{u} (z_{1-}) h_v (0) | \bar B \rangle.      
\end{equation}
This gives ${\cal V}_{9-16}=0$, since for these functions the 
heavy and light quark are in a colour-octet configuration. We conclude
that the tree level contributions are suppressed, and that  we
need a better model for a quantitative estimate. 

To account for the suppression of the colour-octet matrix elements, 
we introduce a parameter $\epsilon$ and write (now including colour) 
\begin{eqnarray} 
&&\langle \bar B | \bar{h}_v (x_-) T^A u (z_{2-})\bar{u} (z_{1-}) T^A h_v(0) 
                | \bar B \rangle
\nonumber \\   
&& \hspace*{2cm}\quad \to  \quad 
  \varepsilon \,\langle \bar B | \bar{h}_v (x_-) u(z_{2-})  | 0 \rangle
              \langle 0 |  \bar{u} (z_{1-}) h_v (0) | \bar B \rangle.      
\label {assume2}
\end{eqnarray}
For local four-quark operators the agreement of calculations of the
lifetime difference of the $B^-$ and $\bar B^0$ 
mesons \cite{Beneke:2002rj,Franco:2002fc} with observations 
suggests that $|\varepsilon|$
cannot be larger than 0.1 without invoking large cancellations between
different terms. Assuming (\ref{assume2}) we can express the
four-quark shape-functions in terms of the square of the $\bar B$ meson 
light-cone distribution amplitude. The resulting parameterizations 
of $v_{s,a}(\omega)$ exhibit a delta-function singularity at 
$\omega=m_b-M_B$, which can be traced to the fact that in the 
vacuum saturation approximation no dynamical (soft) momentum is 
transferred between the $\bar h_v u$ and the 
$\bar u h_v$ configuration. This is clearly not the case in reality,
where soft gluons must be exchanged simply to rearrange colour. 
This motivates our ``modified vacuum saturation approximation,'' 
where we allow the intermediate state to carry soft momentum $k$, 
and write 
\begin{eqnarray} 
\label {model}
 &&\langle \bar B | \bar{h}_v (x_-) T^A u (z_{2-})\bar{u} 
(z_{1-}) T^A h_v(0) | \bar B \rangle
\nonumber \\ && 
\hspace*{2cm}\quad \to 
\varepsilon \int dk_+    f(k_+) 
\langle \bar B | \bar{h}_v (x_-) u(z_{2-})  | k \rangle
              \langle k |  \bar{u} (z_{1-}) h_v (0) | \bar B \rangle,
\end{eqnarray}
where $| k \rangle$ represents a soft state of momentum $k$ and 
$k_+=n_- k$. $f(k_+)$ (whose integral is normalized to 1) 
represents a (non-perturbative) distribution function of the 
$k_+$ momentum component of the soft state. We then write 
\begin{eqnarray}
&& \langle k |\bar{u} (y_-)_\alpha  h_v(x_-)_\beta| \bar B \rangle = 
e^{-i (M_B-m_b-k_+) x_+}\,
\langle k |\bar{u} (y_--x_-)_\alpha  h_v(0)_\beta| \bar B \rangle
\nonumber \\[0.2cm]
&& \hspace*{1cm} \approx \,e^{-i (M_B-m_b-k_+) x_+}\,
\langle 0 |\bar{u} (y_--x_-)_\alpha  h_v(0)_\beta| \bar B \rangle
\nonumber \\[0.2cm]
&& \hspace*{1cm} =\,e^{-i (M_B-m_b-k_+) x_+}\,
\frac{(-i) f_B M_B}{2\sqrt{2 M_B}} \left(
\frac{1+\fmslash{v}}{2} \gamma_5\right)_{\beta\alpha}  
\!\tilde\phi_{B+}(t)  + \mbox{$\slash{n}_-$ - term}
\label{softm}
\end{eqnarray}
with $t=(y-x)_+$. The transition to the second line involves 
the assumption that the $t$-dependence of the matrix element is 
not modified by the ``soft'' state. This may be difficult to justify,
but it allows us to obtain the desired matrix element in terms of the
light-cone distribution amplitude of the 
$\bar B$ meson \cite{Grozin:1996pq,Beneke:2000wa} about which a few things
are known. A second possible 
structure proportional to $\slash{n}_-$ is not given above, because it
drops out in the following calculations. The momentum space
distribution amplitude is defined by 
\begin{equation}
\tilde\phi_{B_+}(t) = \int_0^\infty d \omega \, 
e^{-i\omega t} \,\phi_{B+}(\omega). 
\end{equation}

It is now straightforward to insert (\ref{softm}) into (\ref{model}) 
and to match the Dirac structure to those defining the general
decomposition of the four-quark shape-function. We then find that the
two relevant effective shape-functions are given by 
\begin{eqnarray}
v_s (n_- p)  &=&  - v_a (n_- p) = 
\frac{\pi\alpha_s }{4}\,f_B^2 M_B \,\varepsilon 
\int d k_+\,f(k_+)\Bigg\{\delta(n_- P-k_+) \left[\int_0^\infty
\frac{d\omega}{\omega}\,\phi_{B+}(\omega)\right]^2
\nonumber\\
&& \hspace*{0cm}+\,
\frac{2\phi_{B+}(n_- P-k_+)}{n_- P-k_+} \, 
P\int_0^\infty d\omega\,\frac{\phi_{B+}(\omega)}{n_- P-\omega-k_+}
\Bigg\}
\nonumber\\
&= & 
\frac{\pi\alpha_s }{4}\,\frac{f_B^2 M_B}{\lambda_B^2} \,\varepsilon 
\,f(n_- P)\,
\Bigg\{1+ \frac{2\lambda_B^2}{f(n_- P)} \nonumber\\
&& \hspace*{0cm}
\times \int_0^{n_- P} d k_+ \,f(n_- P-k_+)\,\frac{\phi_{B+}(k_+)}{k_+} \, 
P\int_0^\infty d\omega\,\frac{\phi_{B+}(\omega)}{k_+ - \omega}
\Bigg\},
\label{modfinal}
\end{eqnarray}
where $P$ denotes the principal value and $1/\lambda_B=
\int_0^\infty d\omega \,\phi_{B+}(\omega)/\omega$. 
We observe that the final result is expressed in terms of the 
hadronic variable $n_- P\equiv P_+ = E_H - |\vec{P}_H| $. 
We can get an idea of the magnitude of the power correction 
from the four-quark shape-functions from (\ref{final2}), which 
instructs us to compare $4 v_s(n_- p)/n_+ p$ to 
$S(n_- p)\equiv \hat S(n_- p+M_B-m_b)=\hat S(n_- P)$. The integration 
over $n_+ p$ replaces $n_+ p$ by an average value $\langle n_+ p
\rangle$ 
of order $M_B$, hence
\begin{equation}
\frac{4 v_s(n_- p)}{\langle n_+ p\rangle\hat S(n_- P)} \approx 
\pi\alpha_s\frac{f_B^2}{\lambda_B^2} \,\varepsilon 
\,\frac{M_B}{\langle n_+ p\rangle}
\,\frac{f(n_- P)}{\hat S(n_- P)} \Big\{1+\ldots\Big\},
\end{equation}
where the ellipses in brackets stand for the twofold integral in 
(\ref{modfinal}). If we set the bracket to one and assume that 
$f$ and $\hat S$ have similar shapes, we see that the size of the
correction is about 
\begin{equation}
\pi\alpha_s\frac{f_B^2}{\lambda_B^2} \,\varepsilon 
\approx \pm 5\%,
\label{est}
\end{equation}
compatible with a power correction. Evidently there is a large
uncertainty associated with this estimate, since, besides the
assumptions in the construction of the model, $\alpha_s$ could be
anything between 0.3 (corresponding to the hard-collinear scale) and 1
(corresponding to the soft scale) and $|\varepsilon|=0.1$ is just a 
guess. We should emphasize that the estimate is for the 
semi-leptonic decay of a $B^-$. For $\bar B^0$ the correction vanishes in
our approximation. The distinctive feature of the four-quark
contributions is therefore a difference in the $B^-$ and $\bar B^0$
spectra, which could be of order $5\%$. 

\begin{figure}[t]
   \hskip-1.8cm
   \epsfxsize=12cm
   \centerline{\epsffile{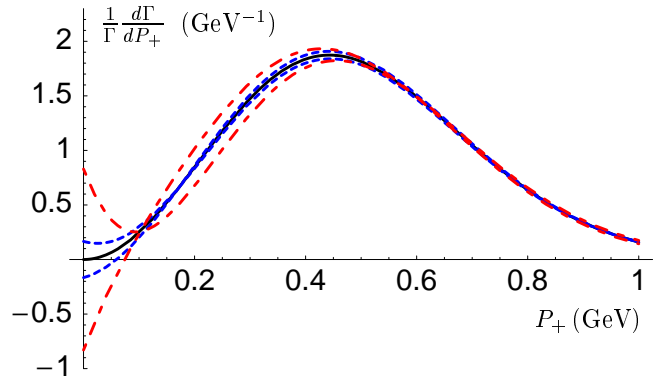}}
   \vspace*{0.0cm}
\centerline{\parbox{14cm}{\caption{\label{fig:plot}\small  
Distortion of the $P_+$ spectrum in $B^-\to X_u\ell \nu$ decay 
by four-quark contributions assuming 
the model (\ref{shapemodel}). The solid central curve is 
$\hat S(P_+)$. The dashed curves correspond to $\lambda=100\,$MeV 
(long dashes) and $\lambda=500\,$MeV (short dashes). Each pair 
is for $\varepsilon=\pm 0.1$.
 }}}
\end{figure}

The effect could be more important than the estimate suggests, if it
leads to a significant distortion of the spectrum in the region of 
$n_- P\sim \Lambda_{\rm QCD}$. For instance, 
if $f(n_- P)$ were peaked at smaller arguments than 
$\hat S(n_- P)$, as would be the case in the naive vacuum saturation 
approximation, where the ``soft'' intermediate state has no momentum, 
the four-quark contribution could give a significant enhancement 
at small values of $n_- P$. To be definite, let us assume 
the following models for the various functions:
\begin{eqnarray}
\hat S(\omega) &=&\frac{32 \omega^2 }{\pi^2 \bar\Lambda^3}
\exp\Big[-\frac{4 \omega^2}{\pi \bar\Lambda^2}\Big] ,
\nonumber\\ 
\phi_{B+} (\omega) &=& \left(\frac{3}{2 \bar\Lambda}\right)^2 
\,\omega\,e^{-3 \omega/ 2 \bar\Lambda} ,
\nonumber\\
f(k_+) &=& \frac{1}{\lambda} \,e^{-k_+ / \lambda}.
\label{shapemodel}
\end{eqnarray} 
To display the effect on a decay spectrum, it is natural to consider
the distribution in the hadronic light-cone variable 
$P_+=E_H - |\vec{P}_H|$, since it is proportional to $\hat S(P_+)$ at
leading power. Including the four-quark shape-functions but 
neglecting the other power corrections, we obtain for 
$B^-\to X_u \ell\nu$
\begin{equation}
\frac{1}{\Gamma} \frac{d  \Gamma}{dP_+} = 
%\frac{G_F^2 |V_{ub}|^2 M_B^5}{192 \pi^3}  
\hat S(P_+) - \frac{10 v_s(n_- p) + 2v_a(n_- p)}{3 m_b}. 
\end{equation}
In Figure~\ref{fig:plot} we show the effect of the four-quark 
contribution on the spectrum of $B^-\to X_u\ell\nu$ decay. 
We choose the parameters $\alpha_s=1$, $f_B = 200\,$MeV, 
$\bar\Lambda = 500$ MeV, $\varepsilon = \pm 0.1$ and plot 
the spectrum for $\lambda = 100\,$MeV and  
$\lambda = 500\,$MeV. The solid curve is the spectrum without 
the power correction, which in the present approximation is also 
the decay spectrum for the neutral $\bar B^0$ meson decay. 
For the smaller value of $\lambda$ the effect can be large, but it 
is concentrated at $P_+<0.5\,$GeV as expected. Clearly, this is 
a rather model-dependent statement. Since the integral of $v_{s,a}(n_- p)$ 
vanishes, the effect changes sign at some $P_+$. Once a sufficiently 
large interval is integrated over, the integrated effect is not expected 
to exceed the estimate (\ref{est}). However, we still expect the 
four-quark contributions to provide the largest difference between 
the spectra of charged and neutral $\bar B$ meson decays. 

%%%%%%%%%%%%%%%%%%%%%%%%%%%%%%%%%%%%%%%%%%%%%%%%%%%%%%%%%%%%%%%%%%%
%%%%%%%%%%%%%%%%%%%%%%%%%%%%%%%%%%%%%%%%%%%%%%%%%%%%%%%%%%%%%%%%%%%

\section{Conclusions}
\label{sec:conclude}

Using the framework of soft-collinear effective theory, 
we have investigated the $1/m_b$ corrections to inclusive heavy-to-light 
transitions in the endpoint region, where the heavy quark 
expansion in local operators breaks down. 
We find that SCET factorizes short- and long-distance effects also at 
sub-leading power. The hadronic tensor, from which all decay spectra are 
derived, is represented as convolutions of hard coefficient functions, 
perturbative jet-functions and non-perturbative shape-functions. 
However, the structure is significantly more complicated than at leading 
power, because the factorization formula involves multiple convolutions 
and many shape-functions. 
We have given the general form of the effective currents and 
shape-functions that can appear at order $1/m_b$.

We then computed the hadronic tensor to order $1/m_b$ in the tree 
approximation. In this approximation the power corrections 
can be parameterized by a few ``effective shape-functions'' of a single 
variable $P_+$ only. Compared to earlier results, we find a new 
contribution from four-quark operators. The other contributions 
agree with previous results for some differential distributions, but 
disagree with others. We estimated the numerical effect of the four-quark 
shape-functions. The effect can lead to a
distortion of the  $P_+$ spectrum of 
$B^- \to X_u \ell \bar{\nu}$ relative to
the one of $\bar B^0 \to X_u \ell \bar{\nu}$. In our model the distortion 
can be significant at small $P_+$ and could be of order a few percent 
when integrated over an interval of order $\Lambda_{\rm QCD}$. 
(The effect vanishes when integrated over all $P_+$.) Our results 
suggest that it should suffice to include the $1/m_b$ corrections 
in the tree approximation in the analysis of measurements. This is 
fortunate, since a prohibitively large number of unknown shape-functions 
is expected to appear once radiative corrections are included.

\vskip0.2cm\noindent
While this paper was being written, two articles 
\cite{Lee:2004ja,Bosch:2004cb} appeared that also 
address $1/m_b$ corrections to $\bar B$ decay distributions in the
soft-collinear effective theory framework. The paper by Lee and Stewart 
\cite{Lee:2004ja} is similar in scope to the present work. Since they
use a different 
representation of SCET, we find it difficult to compare the results 
in detail, but we can nonetheless make several comments. The structure
of factorization as given by our equation (\ref{eq:masfact2})
agrees with their equation (119) on the point
that the most general factorization formula contains convolutions
over additional tri- and tetra-local shape-functions whose 
jet-functions vanish at tree-level.  
However, it appears to us that their basis of 
SCET currents at relative order $\lambda^2$ cannot be complete because 
it does not include four-body operators, which implies that the set
of power-suppressed jet-functions quoted in their equation (119) 
is also incomplete. On the other hand, Lee and Stewart observe that 
$q_s$ in ${\cal L}^{(1)}_{\xi q}$ can be replaced by 
$\slash{n}_+\slash{n}_- q_s/4$, which implies a simplification 
of the general decomposition of the 
four-quark matrix element (\ref{fourquarkshape}). We find that 
only 6 out of the 16 invariant functions defined in (\ref{fourquarkshape}) 
can appear to any order in the 
perturbative expansion. (Two of the 8 functions defined in
\cite{Lee:2004ja} can be eliminated.) 
We differ from \cite{Lee:2004ja} on the 
phenomenological implications of our results. 
On the basis of power counting it is estimated in  \cite{Lee:2004ja}
that the effect of the four-quark contributions is up to 180\% in large 
disagreement with our estimate of 5\% based on a simple model. The 
discrepancy can be attributed to the absence of the colour 
suppression factor $\varepsilon$ in their estimate and numerical 
factors. Bosch, Neubert, and Paz \cite{Bosch:2004cb} 
restrict their analysis to the 
tree approximation, but use the position space SCET formalism as 
we do. Our Section~\ref{sec:tree} has significant overlap with 
their paper and the tree level results are in complete agreement (as is most 
evident by comparing our (\ref{final1}) to their equation (28)).
Our (effective) 
shape-functions can be obtained from theirs by making the replacements
\begin{eqnarray}
S(\omega)&\to& S(-\omega) \nonumber \\
s(\omega)&\to&\frac{s_{\rm kin}(-\omega)
+C_{\rm mag}(m_b/\mu) s_{\rm mag}(-\omega)}{2} \nonumber \\
t(\omega)&\to&t(-\omega) \nonumber \\
\tilde u(\omega)&\to& -u_s(-\omega)-2 v_s(-\omega) \nonumber \\
\tilde v(\omega)&\to& u_a(-\omega)-2 v_a(-\omega). 
\end{eqnarray}
These authors perform a more extensive phenomenological study of 
decay spectra, but they do not consider the four-quark contributions 
explicitly. Rather, they absorb $v_{s,a}$ into a redefinition 
of $u_{s,a}$ (their $\tilde u,\tilde v$). 
While this is technically possible, it is also  
misleading, because it hides the fact that the four-quark contributions 
are different for charged and neutral $\bar B$ meson decay. In a recent 
paper \cite{Neubert:2004cu} Neubert estimated the numerical 
effect of the four-quark shape-functions in various models, including
the naive vacuum saturation ansatz ($f(k_+)=\delta(k_+))$. 
When integrated over a
sufficiently large region in $P_+$, his result is in qualitative
agreement with ours.

\vskip0.4cm\noindent
{\bf Acknowledgements}

\vskip0.2cm\noindent 
This work was supported by the DFG 
Sonderforschungsbereich/Transregio 9 ``Computer-gest\"utzte Theoretische 
Teilchenphysik''. M.B. would like to thank the KITP, Santa Barbara, 
and the INT, Seattle for their generous hospitality while part 
of the work was being done. F.C. acknowledges support of the 
DFG Graduiertenkolleg ``Hochenergiephysik und Astroteilchenphysik''.

\end{document}